\documentclass[11pt,reqno]{amsart}

\usepackage{amsfonts}
\usepackage{mathrsfs}
\allowdisplaybreaks[4]
\usepackage[dvips]{graphicx} 
\usepackage{epsfig}
\usepackage{amssymb}
\usepackage{epsfig}
\usepackage{amsmath}
\usepackage{cite}

\newcommand{\be}{\begin{equation}}
\newcommand{\ee}{\end{equation}}
\newcommand{\bea}{\begin{eqnarray}}
\newcommand{\eea}{\end{eqnarray}}
\newcommand{\ba}{\begin{array}}
\newcommand{\ea}{\end{array}}
\newcommand{\bean}{\begin{eqnarray*}}
\newcommand{\eean}{\end{eqnarray*}}

\newcommand{\pa}{\partial}

\begin{document}

\title[Darboux]{ The n-order rogue waves of Fokas-Lenells equation}
\author{Shuwei Xu $^{1}$, Jingsong He $^{2}$ $^*$,  Yi Cheng $^{1}$
and  K. Porseizan $^{3}$
 }

\thanks{$^*$ Corresponding author: hejingsong@nbu.edu.cn,jshe@ustc.edu.cn}

 \maketitle
\dedicatory {$^{1}$School of Mathematical Sciences, USTC, Hefei,
Anhui 230026, P.\ R.\ China\\
\mbox{\hspace{0.75cm}}$^{2}$Department of Mathematics, Ningbo
University, Ningbo, Zhejiang
315211, P.\ R.\ China\\
\mbox{\hspace{0.75cm}}$^{3}$Department of Physics, Pondicherry
University, Puducherry 605014, India }

\begin{abstract}
Considering certain terms of the next asymptotic order beyond the nonlinear Schr\"odinger equation (NLS) equation,
the Fokas-Lenells (FL) equation governed by the FL system arise as a
model for nonlinear pulse propagation in optical fibers. The
expressions of the $q^{[n]}$ and $r^{[n]}$ in the FL system are
generated by n-fold Darboux transformation (DT), each element of the
matrix is a 2¡Á2 matrix, expressed by a ratio of
$(2n+1)\times (2n+1)$ determinant and $2n\times 2n$ determinant of
eigenfunctions. Further, a Taylor series expansion about the
n-order breather solutions $q^{[n]}$ generated using DT by assuming
periodic seed solutions under reduction can generate the n-order rogue
waves of the FL equation explicitly with $2n+3$ free parameters.
\end{abstract}

\mbox{\hspace{3cm}} Darboux transformation, breather solution, rogue
wave.

{\bf PACS(2010) numbers}: 02.30.Ik, 42.81.Dp, 52.35.Bj, 52.35.Sb, 94.05.Fg


\section{introduction}

The Fokas-Lenells equation
(FL)\cite{Fokas1,Lenells1,Lenells2,Lenells4,Lenells3}
\begin{equation}\label{FL}
iq_{xt}-iq_{xx}+2q_{x}-q_{x}qq^{\ast}+iq=0,
\end{equation}
is one of the important models from both mathematical and  physical
considerations. In eq.(\ref{FL}), $q$ represents the complex field
envelope and asterisk denotes its complex conjugation, the subscript $x$
(or $t$) denotes partial derivative with respect to $x$ (or $t$).
The FL equation \cite{Fokas1,Lenells2} is related to the nonlinear
Schr\"odinger (NLS) equation in the same way as the Camassa-Holm
equation associated with the KdV equation. The authors of
\cite{Lenells2} , after deriving associated Lax pair and using
initial value problem, they were able to solve the equation.  The
soliton solutions of the FL equation have been constructed via the
Riemann-Hilbert method in \cite{Lenells1} and through dressing method in
\cite{Lenells4}.  The breather solutions of the FL equation have also
been constructed via a dressing-Backlund transformation related to
the Riemann-Hilbert problem formulation of the inverse scattering
theory \cite{Wright}; also in the same paper three instability
regions were analyzed, associated with a single unstable wave
number.  The FL equation actually describes  the first negative flow
of the integrable hierarchy associated with the derivative nonlinear
Schr\"odinger (DNLS) equation \cite{Lenells3} (it also belongs to
the deformed DNLS hierarchy proposed by A.Kundu \cite{Kundu1,Kundu2}). The
lattice representation and the dark solitons of the FL equation have
been presented in \cite{Vekslerchik}, where a relationship is also
established between the FL equation and other integrable models
including the NLS equation, the Merola-Ragnisco-Tu equations and the
Ablowitz-Ladik equation. Recently, it has been shown that the
periodic initial value problem for the FL equation is well-posed in
a Sobolev space with exponent greater than 3/2\cite{Fokas2}. In
optics, considering suitable higher order linear and nonlinear
optical effects,  the FL equation has been derived as a model to
describe the femtosecond pulse propagation through single mode optical
silica fiber and several interesting solutions have been constructed
\cite{Lenells3}.

Rogue waves have recently been the subject of intensive
investigations in oceanography\cite{Kharif1,Kharif2,Akhmediev2 },
where they occur due to either modulation instability
\cite{Peregrine,Zakharov1,Akhmediev3}, or other
processes\cite{Didenkulova,Ying,lai}. The first order rogue wave
usually takes the form of a single peak hump with two caves in a
plane with a nonzero boundary. One of the possible generating
mechanisms for rogue waves is through the creation of breathers
which can be formed as a consequence of modulation instability.  Then, larger
rogue waves can emerge when two or more breathers collide with each
other\cite{Akhmediev4,limits,hefokas1}. Rogue waves have also been
observed in space plasmas\cite{ruderman,derman,Moslem,xuhe,xxuhe},
as well as in optics when propagating high power optical radiation
through photonic crystal fibers \cite{Optical1,Optical2,Optical3}.
Though the rogue waves have been reported in different branches of
physics where the system dynamics is governed by single equation, to
the best our knowledge, they have been observed and reported very
little in the coupled systems. For example, Rogue waves of the
coupled Schr\"odinger equations are construct in the
literature\cite{GuoBL2,Baronio,WeiGuo}. In experiment, the rogue
waves in a multistable system \cite{Rider} is revealed by
experiments with an erbium-doped fiber laser driven by harmonic pump
modulation. Considering higher order effects in the propagation of
femtosecond pulses, rogue waves have been reported in the Hirota
equation\cite{Akhmediev5,taohe1} and in resonant erbium-doped fibre
system governed by a coupled system of the nonlinear Schr\"odinger
equation and the Maxwell-Bloch (NLS-MB) equations\cite{NewTypes}.
Very recently, the new types of  matter rogue waves \cite{Zhenyun}
have been reported in $F=1$ spinor Bose-Einstein condensate governed by a
three-component NLS equations. Some  interesting results on the
multi-rogue wave solutions of the NLS equation have also been done in
\cite{Dubard2,Gaillard,triplets,guo,Yang,Circular,Modulation}, which
shows that there exist many patterns of the rogue waves and their
formulae have extreme complexity.\\

Considering the physical significance of the FL equation,
inspired by the importance of the recent interesting developments in
the rogue waves of the NLS type equations, so we have  reported  the
first order rogue wave\cite{jhswk} of the FL equation in  by Darboux
transformation (DT)\cite{GN,matveev,he}. Our construction reveals
that there exists some deviations in their solutions and DT between
the FL system and other integrable models such as
Ablowitz-Kaup-Newell-Segur (AKNS) system\cite{AKNS,Ablowitz} and
Kaup-Newell (KN) system\cite{KN,kenji1,steduel}. The purpose of this
paper is to provide a detailed derivation of the determinant
representation of the N-fold DT of the FL equation as we have done
for the case of NLS equation \cite{he}, which will be used to construct higher
order rogue wave. Several different patterns of the higher order
rogue waves will be plotted according to the determinant
representation.\\

The organization of this paper is as follows. In section 2, it provides a relatively
simple approach to DT for the FL system, which is followed by the determinant
representation of the n-fold DT and formulae of $q^{[n]}$ and
$r^{[n]}$ expressed by eigenfunctions of spectral problem are given.
The reduction of DT of the FL system to the FL equation is also
discussed by choosing paired eigenvalues and eigenfunctions. In
section 3, a Taylor series expansion about the n-order breather
solutions  generated by  DT from a periodic seed solution with a
constant amplitude can construct the  n-order rogue waves of the FL
equation in the determinant forms with $2n+3$ free parameters.
Finally, we conclude the results in section 4.

\section{Darboux transformation}

Let us start from the non-trivial flow of the FL (Fokas-Lenells) system\cite{Lenells2},
\begin{equation}\label{sy1}
iq_{xt}-iq_{xx}+2q_{x}-q_{x}qr+iq=0,
\end{equation}
\begin{equation}\label{sy2}
ir_{xt}-ir_{xx}-2r_{x}+r_{x}rq+ir=0,
\end{equation}
which are  exactly reduced to the FL eq.(\ref{FL}) for $r=q^\ast$ while the choice $r
=-q^\ast$ would lead to eq.(\ref{FL}) with the sign of the nonlinear term changed. The Lax pairs
corresponding to coupled FL eq.(\ref{sy1}) and (\ref{sy2}) can
be given by the FL spectral problem\cite{Lenells2}
\begin{equation}\label{sys11}
 \pa_{x}\psi=(J\lambda^2+Q\lambda)\psi=U\psi,
\end{equation}
\begin{equation}\label{sys22}
\pa_{t}\psi=(J\lambda^2+Q\lambda+V_{0}+V_{-1}\lambda^{-1}+\dfrac{1}{4}J\lambda^{-2})\psi=V\psi,
\end{equation}
with
\begin{equation}\label{fj1}
    \psi=\left( \begin{array}{c}
      \phi \\
      \varphi\\
     \end{array} \right),\nonumber\\
   \quad J= \left( \begin{array}{cc}
      -i &0 \\
      0 &i\\
   \end{array} \right),\nonumber\\
  \quad Q=\left( \begin{array}{cc}
     0 &q_{x} \\
     r_{x} &0\\
  \end{array} \right),\nonumber\\
\end{equation}
\begin{equation}\label{fj2}
V_{0}=\left( \begin{array}{cc}
i-\frac{1}{2}iqr &0 \\
 0&-i+\frac{1}{2}iqr\\
\end{array} \right),\quad V_{-1}=\left( \begin{array}{cc}
0 &\frac{1}{2}iq \\
- \frac{1}{2}ir&0\\
\end{array} \right).\nonumber\\
\end{equation}
Here $\lambda$ is an arbitrary complex number, called the eigenvalue(or spectral parameter),
and $\psi$ is called the eigenfunction associated with  $\lambda$ of the FL system.
Equations (\ref{sy1}) and (\ref{sy2}) are equivalent to the integrability condition  $U_{t}-V_{x}+[U,V]=0$ of
(\ref{sys11}) and (\ref{sys22}).\\

The main task of this section is to present a detailed derivation of
the DT of the FL equation and the determinant
representation of the n-fold transformation. Based on the DT for the
NLS\cite{GN,matveev,he} and the
DNLS\cite{kenji1,steduel,xuhe,xxuhe},
 the main steps are : 1) finding $2\times 2$ matrix  $T$ so that the FL spectral problem
 eq.(\ref{sys11}) and eq.(\ref{sys22}) is covariant, then get new
 solution $(q^{[1]},r^{[1]})$ expressed by elements of $T$ and
 seed solution $(q,r)$; 2) finding the expressions of elements of $T$ in terms of eigenfunctions of FL spectral problem
corresponding to the seed solution $(q,r)$; 3) to get the determinant representation of n-fold DT $T_n $ and
new solutions $(q^{[n]},r^{[n]})$ by  $n$-times iteration of the DT;
4) to consider the reduction condition: $q^{[n]}= (r^{[n]})^*$ by choosing special eigenvalue $\lambda_k$
 and its eigenfunction $\psi_k$, and then get $q^{[n]}$ of the FL equation expressed by its seed solution
 $q$ and its associated eigenfunctions $\{\psi_k,k=1,2,\cdots,n\}$. However, we shall use the kernel of n-fold DT($T_n$) to
fix it in the third step instead of iteration.

It is straightforward to see that the spectral problem (\ref{sys11}) and (\ref{sys22}) are transformed to
\begin{equation}\label{bh1}
{\psi^{[1]}}_{x}=U^{[1]}~\psi^{[1]},\ \ U^{[1]}=(T_{x}+T~U)T^{-1}.
\end{equation}
\begin{equation}\label{bh2}
{\psi^{[1]}}_{t}=V^{[1]}~\psi^{[1]}, \ \ V^{[1]}=(T_{t}+T~V)T^{-1}.
\end{equation}\\
under a gauge transformation \begin{equation}\label{bh3}
\psi^{[1]}=T~\psi.
\end{equation}
By cross differentiating (\ref{bh1}) and (\ref{bh2}), we obtain
\begin{equation}\label{bh4}
{U^{[1]}}_{t}-{V^{[1]}}_{x}+[{U^{[1]}},{V^{[1]}}]=T(U_{t}-V_{x}+[U,V])T^{-1}.
\end{equation}
This implies that, in order to make eqs.(\ref{sy1}) and (\ref{sy2}) invariant under the
transformation (\ref{bh3}), it is crucial to search a matrix $T$ so that  $U^{[1]}$, $V^{[1]}$ have
the same forms as $U$, $V$. At the same time the old potential(or
seed solution)($q$, $r$) in spectral matrixes $U$, $V$ are mapped
into new potentials(or new solution)($q^{[1]}$, $r^{[1]}$) in transformed spectral matrixes $U^{[1]}$, $V^{[1]}$.\\

2.1 One-fold Darboux transformation of the FL system\\

Considering the universality of DT, suppose that the trial Darboux matrix $T$ in eq.(\ref{bh3}) is of the
form
\begin{equation}\label{tt1}
T=T(\lambda)=\left( \begin{array}{cc}
a_{1}&b_{1} \\
c_{1} &d_{1}\\
\end{array} \right)\lambda+\left( \begin{array}{cc}
a_{0}&b_{0}\\
c_{0} &d_{0}\\
\end{array} \right)+\left( \begin{array}{cc}
a_{-1}&b_{-1}\\
c_{-1} &d_{-1}\\
\end{array} \right)\lambda^{-1},
\end{equation}
where $a_{-1},  b_{-1},  c_{-1},   d_{-1},a_{0},  b_{0},  c_{0},   d_{0},   a_{1},   b_{1},   c_{1},   d_{1}$ are functions of $x$ and $t$. From \begin{equation}\label{tt2} T_{x}+T~U=U^{[1]}~T,
\end{equation}
comparing the coefficients of $\lambda^{j}, j=3, 2, 1, 0,-1$, it yields
\begin{eqnarray}\label{xx1}
&&\lambda^{3}: b_{1}=0,\ c_{1}=0,\nonumber\\
&&\lambda^{2}: q_{x}~a_{1}+2~i~b_{0}-{q^{[1]}}_{x}d_{1}=0,\ -{r^{[1]}}_{x}~a_{1}+r_{x}~d_{1}-2~i~c_{0}=0,\nonumber\\
&&\lambda^{1}: {a_{1}}_{x}+r_x~b_{0}-{q^{[1]}}_{x}c_{0}=0,\ {d_{1}}_{x}+q_x c_{0}-{r^{[1]}}_{x}b_{0}=0,\nonumber\\
&&\mbox{\hspace{0.7cm}} q_{x} a_{0}-{q^{[1]}}_{x}d_{0}+2ib_{-1}=0,-{r^{[1]}}_{x}a_{0}+r_{x} d_{0}-2ic_{-1}=0,\nonumber\\
&&\lambda^{0}:-q^{[1]}c_{-1}+r_{x}b_{-1}+{a_{0}}_{x}=0,-q^{[1]}d_{-1}+q_{x}a_{-1}+{b_{0}}_{x}=0,\nonumber\\
&&\mbox{\hspace{0.7cm}} -r^{[1]}a_{-1}+r_{x}d_{-1}+{c_{0}}_{x}=0,-r^{[1]}b_{-1}+q_{x}c_{-1}+{d_{0}}_{x}=0,\nonumber\\
&&\lambda^{-1}: {a_{-1}}_{x}={b_{-1}}_{x}={c_{-1}}_{x}={d_{-1}}_{x}=0.
\end{eqnarray}
The terms in the previous equation $a_{-1},b_{-1},c_{-1},d_{-1} $ are all functions of $t$ only.
Similarly, from  \begin{equation}\label{tt3} T_{t}+T~V=V^{[1]}~T,
\end{equation} comparing the coefficients of $\lambda^{j} ,j =2, 1, 0,-1,-2,-3$ under the condition $b_{1}=c_{1}=0$, it implies
\begin{eqnarray}\label{ttt1}
&&\lambda^{-3}: b_{-1}=c_{-1}=0,\nonumber\\
&&\lambda^{-2}: -q^{[1]}d_{-1}+q a_{-1}+b_{0}=0,-r^{[1]}a_{-1}+r d_{-1}+c_{0}=0,\nonumber\\
&&\lambda^{-1}: {a_{-1}}_{t}-\dfrac{1}{2}iq^{[1]}c_{0}+\frac{1}{2}iq^{[1]}r^{[1]}a_{-1}-\frac{1}{2}iqra_{-1}-\frac{1}{2}irb_{0}=0,-q^{[1]}d_{0}+q a_{0}=0,\nonumber\\
&&\mbox{\hspace{0.8cm}}{d_{-1}}_{t}+\dfrac{1}{2}ir^{[1]}b_{0}-\frac{1}{2}iq^{[1]}r^{[1]}d_{-1}+\frac{1}{2}iqrd_{-1}+\frac{1}{2}iqc_{0}=0,-r^{[1]}a_{0}+r d_{0}=0,\nonumber\\
&&\lambda^{0}: {a_{0}}_{t}-\dfrac{1}{2}iqra_{0}+\frac{1}{2}iq^{[1]}r^{[1]}a_{0}=0,{d_{0}}_{t}+\dfrac{1}{2}iqrd_{0}-\frac{1}{2}iq^{[1]}r^{[1]}d_{0}=0,\nonumber\\
&&\mbox{\hspace{0.8cm}}q_{x}a_{-1}+\frac{1}{2}iqa_{1}-2ib_{0}-\frac{1}{2}iq^{[1]}d_{1}+{b_{0}}_{t}-{q^{[1]}}_{x}d_{-1}+\frac{1}{2}iq^{[1]}r^{[1]}b_{0}+\frac{1}{2}iqrb_{0}=0,\nonumber\\
&&\mbox{\hspace{0.8cm}}r_{x}d_{-1}-\frac{1}{2}ird_{1}+2ic_{0}+\frac{1}{2}ir^{[1]}a_{1}+{c_{0}}_{t}-{r^{[1]}}_{x}a_{-1}-\frac{1}{2}iq^{[1]}r^{[1]}c_{0}-\frac{1}{2}iqrc_{0}=0,\nonumber\\
&&\lambda^{1}: -{q^{[1]}}_{x}d_{0}+q_{x}a_{0}=0,-{r^{[1]}}_{x}a_{0}+r_{x}d_{0}=0,\nonumber\\
&&\mbox{\hspace{0.8cm}}\dfrac{1}{2}iq^{[1]}r^{[1]}a_{1}-\frac{1}{2}iqra_{1}+{a_{1}}_{t}+r_{x}b_{0}-{q^{[1]}}_{x}c_{0}=0,\nonumber\\
&&\mbox{\hspace{0.6cm}}-\dfrac{1}{2}iq^{[1]}r^{[1]}d_{1}+\frac{1}{2}iqrd_{1}+{d_{1}}_{t}+q_{x}c_{0}-{r^{[1]}}_{x}b_{0}=0,\nonumber\\
&&\lambda^{2}:q_{x}a_{1}-q^{[1]}_{x}d_{1}+2ib_{0}=0,r_{x}d_{1}-q^{[1]}_{x}a_{1}-2ic_{0}=0.
\end{eqnarray}

In order to get the non-trivial solutions, we shall construct a
basic (or one-fold) DT matrix $T$ under an
assumption that $a_{0} = 0$ and $d_{0} = 0$. If we set $ a_{0} \not=0$, then we can get that
$d_{0}$ is not zero by $-q^{[1]}d_{0}+q a_{0}=0$ and $-q^{[1]}d_{0}+q a_{0}=0$ from eq.(\ref{ttt1}),
 and furthermore find that some coefficients of $a_{0}$ and $d_{0}$  are constants by taking $q^{[1]}=q\dfrac{a_{0}}{d_{0}}$ and $r^{[1]}=r \dfrac{d_{0}}{a_{0}}$ into eq.(\ref{xx1}) and eq.(\ref{ttt1}), which gives a trivial solution. What is more, we can learn those equation $q_{x}~a_{1}+2~i~b_{0}-{q^{[1]}}_{x}d_{1}=0$, $-{r^{[1]}}_{x}~a_{1}+r_{x}~d_{1}-2~i~c_{0}=0$,$-q^{[1]}d_{-1}+q a_{-1}+b_{0}=0$ and $-r^{[1]}a_{-1}+r d_{-1}+c_{0}=0$ from eq.(\ref{xx1}) and eq.(\ref{ttt1}).
 Under the condition $a_{1}d_{1}a_{-1}d_{-1}\neq0$,we can get that $a_{-1}$ , $d_{-1}$,  $a_{1}$ and $d_{1}$  are constants from eq.(\ref{xx1}) and eq.(\ref{ttt1}). This means that our assumption does not suppresses the generality of the DT of the FL system. Based on
 eq.(\ref{xx1}) and (\ref{ttt1}) and above analysis,
let  Darboux matrix $T$ be  the  form  of
\begin{equation}\label{TT}
 T_{1}=T_{1}(\lambda;\lambda_1;\lambda_2)=\left( \begin{array}{cc}
a_{1}&0\\
0 &d_{1}\\
\end{array} \right)\lambda+\left( \begin{array}{cc}
0&b_{0}\\
c_{0} &0\\
\end{array} \right)+\left( \begin{array}{cc}
a_{-1}&0\\
0&d_{-1}\\
\end{array} \right)\lambda^{-1}.
\end{equation}
Here $a_{1}, d_{1},b_{0},c_{0}$ are undetermined function of ($x$, $t$), which will be expressed by the
eigenfunction associated with $\lambda_1$ and $\lambda_2$ in the FL spectral problem, and $a_{-1}, d_{-1}$ are constants.
First of all, we introduce  $n$ eigenfunctions $\psi_j$ as
\begin{eqnarray}
&&\psi_{j}=\left(
\begin{array}{c}\label{jie2}
 \phi_{j}   \\
 \varphi_{j}  \\
\end{array} \right),\ \ j=1,2,....n,\phi_{j}=\phi_{j}(x,t,\lambda_{j}), \
\varphi_{j}=\varphi_{j}(x,t,\lambda_{j}). \label{jie1}
\end{eqnarray}

\noindent {\bf Theorem 1.}{\sl  The  elements of one-fold DT are parameterized by the
eigenfunction $\psi_i$ ($i=1,2$) associated with $\lambda_i$, as follows
 \begin{eqnarray}
T_1(\lambda;\lambda_1;\lambda_2)=\left(
\begin{array}{cc}
a_{1}\lambda +a_{-1}\lambda^{-1}& b_{0}\\
c_{0} & d_{1}\lambda +d_{-1}\lambda^{-1}
\end{array}  \right),  \label{DT1matrix}
\end{eqnarray}
and then the new solutions $q^{[1]}$ and $r^{[1]}$ are given by
\begin{eqnarray}\label{sTT}
q^{[1]}=q \dfrac{a_{-1}}{d_{-1}}+\dfrac{b_{0}}{d_{-1}},
r^{[1]}=r \dfrac{d_{-1}}{a_{-1}}+\dfrac{c_{0}}{a_{-1}},
\end{eqnarray}
with $a_{-1}=1,d_{-1}=1$ and
\begin{eqnarray}
a_{1}=
\dfrac{\begin{vmatrix}
-{\lambda_{1}}^{-1}\phi_{1}&\varphi_{1}\\
-{\lambda_{2}}^{-1}\phi_{2}&\varphi_{2}\nonumber\\
\end{vmatrix}}{\begin{vmatrix}
\lambda_{1}\phi_{1}&\varphi_{1}\\
\lambda_{2}\phi_{2}&\varphi_{2}\nonumber\\
\end{vmatrix}},\quad
d_{1}=
\dfrac{\begin{vmatrix}
-{\lambda_{1}}^{-1}\varphi_{1}&\phi_{1}\\
-{\lambda_{2}}^{-1}\varphi_{2}&\phi_{2}\nonumber\\
\end{vmatrix}}{\begin{vmatrix}
\lambda_{1}\varphi_{1}&\phi_{1}\\
\lambda_{2}\varphi_{2}&\phi_{2}\nonumber\\
\end{vmatrix}},\quad
b_{0}=
\dfrac{\begin{vmatrix}
\lambda_{1}\phi_{1}&-{\lambda_{1}}^{-1}\phi_{1}\\
\lambda_{2}\phi_{2}&-{\lambda_{2}}^{-1}\phi_{2}\nonumber\\
\end{vmatrix}}{\begin{vmatrix}
\lambda_{1}\phi_{1}&\varphi_{1}\\
\lambda_{2}\phi_{2}&\varphi_{2}\nonumber\\
\end{vmatrix}},\quad
c_{0}=
\dfrac{\begin{vmatrix}
\lambda_{1}\varphi_{1}&-{\lambda_{1}}^{-1}\varphi_{1}\\
\lambda_{2}\varphi_{2}&-{\lambda_{2}}^{-1}\varphi_{2}\nonumber\\
\end{vmatrix}}{\begin{vmatrix}
\lambda_{1}\varphi_{1}&\phi_{1}\\
\lambda_{2}\varphi_{2}&\phi_{2}\nonumber\\
\end{vmatrix}}
\end{eqnarray}
and the new eigenfunction $\psi_j^{[1]}$ corresponding to $\lambda_j$ is
\begin{equation}
\psi^{[1]}_j=
\left(
\begin{array}{c}
 \dfrac{\begin{vmatrix}
\lambda_{i}\phi_{i}&-{\lambda_{i}}^{-1}\phi_{i}&\varphi_{i}\\
\lambda_{1}\phi_{1}&-{\lambda_{1}}^{-1}\phi_{1}&\varphi_{1}\\
\lambda_{2}\phi_{2}&-{\lambda_{2}}^{-1}\phi_{2}&\varphi_{2}\nonumber\\
\end{vmatrix}}{\begin{vmatrix}
\lambda_{1}\phi_{1}&\varphi_{1}\\
\lambda_{2}\phi_{2}&\varphi_{2}\nonumber\\
\end{vmatrix}}\\ \\
\dfrac{\begin{vmatrix}
\lambda_{i}\varphi_{i}&-{\lambda_{i}}^{-1}\varphi_{i}&\phi_{i}\\
\lambda_{1}\varphi_{1}&-{\lambda_{1}}^{-1}\varphi_{1}&\phi_{1}\\
\lambda_{2}\varphi_{2}&-{\lambda_{2}}^{-1}\varphi_{2}&\phi_{2}\nonumber\\
\end{vmatrix}}{\begin{vmatrix}
\lambda_{1}\varphi_{1}&\phi_{1}\\
\lambda_{2}\varphi_{2}&\phi_{2}\nonumber\\
\end{vmatrix}}
\end{array}
\right).
\end{equation}
} {\bf Proof:} Note that $a_{-1}$ and $d_{-1}$ are constants, which
are derived from the eq.(\ref{xx1}) and  eq.(\ref{ttt1}),  respectively. From
eq.(\ref{ttt1}) and transformation eq.(\ref{TT}), new solutions are
given by
 \begin{eqnarray} \label{TT1}
q^{[1]}=q \dfrac{a_{-1}}{d_{-1}}+\dfrac{b_{0}}{d_{-1}},
r^{[1]}=r \dfrac{d_{-1}}{a_{-1}}+\dfrac{c_{0}}{a_{-1}}
\end{eqnarray}
By using a general fact of the DT, (i.e.),
 $T_{1}(\lambda;\lambda_{1};\lambda_{2})|_{\lambda=\lambda_i}\psi_{i}=0,i=1,2$, new solutions are given as eq. (\ref{sTT}). Further, by using the explicit matrix
representation eq.(\ref{DT1matrix}) of $T_1$, then
$\psi^{[1]}_j$ is given by
\begin{eqnarray}
&&\psi^{[1]}_j=T_1(\lambda;\lambda_{1};\lambda_{2})|_{\lambda=\lambda_j} \psi_j=\left.\left(
\begin{array}{cc}
\dfrac{\begin{vmatrix}
-{\lambda_{1}}^{-1}\phi_{1}&\varphi_{1}\\
-{\lambda_{2}}^{-1}\phi_{2}&\varphi_{2}\nonumber\\
\end{vmatrix}}{\begin{vmatrix}
\lambda_{1}\phi_{1}&\varphi_{1}\\
\lambda_{2}\phi_{2}&\varphi_{2}\nonumber\\
\end{vmatrix}}\lambda+\lambda^{-1}& \dfrac{\begin{vmatrix}
\lambda_{1}\phi_{1}&-{\lambda_{1}}^{-1}\phi_{1}\\
\lambda_{2}\phi_{2}&-{\lambda_{2}}^{-1}\phi_{2}\nonumber\\
\end{vmatrix}}{\begin{vmatrix}
\lambda_{1}\phi_{1}&\varphi_{1}\\
\lambda_{2}\phi_{2}&\varphi_{2}\nonumber\\
\end{vmatrix}}\\
\dfrac{\begin{vmatrix}
\lambda_{1}\varphi_{1}&-{\lambda_{1}}^{-1}\varphi_{1}\\
\lambda_{2}\varphi_{2}&-{\lambda_{2}}^{-1}\varphi_{2}\nonumber\\
\end{vmatrix}}{\begin{vmatrix}
\lambda_{1}\varphi_{1}&\phi_{1}\\
\lambda_{2}\varphi_{2}&\phi_{2}\nonumber\\
\end{vmatrix}}&\dfrac{\begin{vmatrix}
-{\lambda_{1}}^{-1}\varphi_{1}&\phi_{1}\\
-{\lambda_{2}}^{-1}\varphi_{2}&\phi_{2}\nonumber\\
\end{vmatrix}}{\begin{vmatrix}
\lambda_{1}\varphi_{1}&\phi_{1}\\
\lambda_{2}\varphi_{2}&\phi_{2}\nonumber\\
\end{vmatrix}}\lambda+\lambda^{-1}
\end{array}  \right)\right|_{\lambda=\lambda_j} \left( \begin{array} {c}
\phi_j\\
\varphi_j
\end{array} \right)\\
&&\mbox{\hspace{4.5cm}}=\left(
\begin{array}{c}
 \dfrac{\begin{vmatrix}
\lambda_{i}\phi_{i}&-{\lambda_{i}}^{-1}\phi_{i}&\varphi_{i}\\
\lambda_{1}\phi_{1}&-{\lambda_{1}}^{-1}\phi_{1}&\varphi_{1}\\
\lambda_{2}\phi_{2}&-{\lambda_{2}}^{-1}\phi_{2}&\varphi_{2}\nonumber\\
\end{vmatrix}}{\begin{vmatrix}
\lambda_{1}\phi_{1}&\varphi_{1}\\
\lambda_{2}\phi_{2}&\varphi_{2}\nonumber\\
\end{vmatrix}}\\ \\
\dfrac{\begin{vmatrix}
\lambda_{i}\varphi_{i}&-{\lambda_{i}}^{-1}\varphi_{i}&\phi_{i}\\
\lambda_{1}\varphi_{1}&-{\lambda_{1}}^{-1}\varphi_{1}&\phi_{1}\\
\lambda_{2}\varphi_{2}&-{\lambda_{2}}^{-1}\varphi_{2}&\phi_{2}\nonumber\\
\end{vmatrix}}{\begin{vmatrix}
\lambda_{1}\varphi_{1}&\phi_{1}\\
\lambda_{2}\varphi_{2}&\phi_{2}\nonumber\\
\end{vmatrix}}
\end{array}
\right).
\end{eqnarray}
After, a tedious calculation shows that $T_1$ in eq.(\ref{DT1matrix})
and new solutions indeed satisfy eq.(\ref{tt2}) and eq.(\ref{tt3})
or (equivalently  eq.(\ref{xx1}) and eq.(\ref{ttt1})). So FL
spectral  problem  is  covariant under transformation $T_1$ in
eq.(\ref{DT1matrix}), and thus it is the DT of eq.(\ref{sy1}) and
eq.(\ref{sy2}). $\square$ The remaining issue is how to guarantee
the validity of the reduction condition, i.e.,
$q^{[1]}=(r^{[1]})^{*}$. We shall solve it at the end of this
section by choosing special eigenfunctions and eigenvalues.\\

2.2 N-fold Darboux transformation for FL system\\

The key task is to establish the determinant representation of the n-fold DT
for FL system in this subsection. For this purpose, we set
$$
\begin{array}{cc}
\textbf{D}=&\left\{\left.\left(\begin{array}{cc}
a& 0\\
0&d
  \end{array} \right)\right| a,d \text{ are complex functions of}\ x\ \text{and}\ t  \right\},\\
\textbf{A}=&\left\{\left.\left(\begin{array}{cc}
0& b\\
c&0
  \end{array} \right)\right| b,c \text{ are complex functions of}\ x\ \text{and}\ t  \right\},
\end{array}
$$
as in ref.\cite{kenji1}.

 According to the form of $T_1$ in eq.(\ref{TT}), the n-fold DT should be of the form \cite{kenji1}
\begin{equation}\label{tnss}T_{n}=T_{n}(\lambda;\lambda_1,\lambda_2, \cdots,\lambda_{2n})
=\sum_{l=-n}^{n}P_{l}\lambda^{l},
\end{equation}
with \begin{eqnarray}
\label{tnsss}
P_{n}=\left( \mbox{\hspace{-0.2cm}}
\begin{array}{cc}
a_{n}\mbox{\hspace{-0.3cm}}&0 \\
0 \mbox{\hspace{-0.3cm}}&d_{n}\\
\end{array}  \mbox{\hspace{-0.2cm}}\right)\in \textbf{D},\ P_{n-1}=\left( \mbox{\hspace{-0.2cm}}
\begin{array}{cc}
0 \mbox{\hspace{-0.3cm}}&b_{n-1} \\
c_{n-1} \mbox{\hspace{-0.3cm}}&0\\
\end{array}  \mbox{\hspace{-0.2cm}}\right)\in \textbf{A},\ P_{l}\in \textbf{D}&\textrm{\small \mbox{\hspace{-0.3cm}}(if $l-n$ is even)}
,\ P_{l}\in \textbf{A}&\textrm{\small \mbox{\hspace{-0.3cm}}(if $l-n$ is odd)}.
\end{eqnarray}
Here $P_{-n}$ is a constant matrix with $a_{-n}=d_{-n}=1$, $P_i(-(n-1)\leq i\leq n)$ is the function of $x$ and $t$.
Specifically, from algebraic equations,
\begin{equation}\label{ttnss}
\psi_{k}^{[n]}=T_{n}(\lambda;\lambda_1,\cdots,\lambda_{2n})|_{\lambda=\lambda_k}\psi_{k}=\sum_{l=-n}^{n}P_{l}\lambda_{k}^{l}\psi_{k}=0,
k=1,2,\cdots,2n,
\end{equation}
coefficients $P_i$ is solved by Cramer's rule. Thus we get determinant representation of the $T_n$.

\noindent {\bf Theorem2.} The n-fold DT of the FL system can be expressed by
\begin{equation}
\label{fss1}T_{n}=T_{n}(\lambda;\lambda_1,\lambda_2,\cdots,\lambda_{2n})=\left(
\begin{array}{cc}
\dfrac{\widetilde{(T_{n})_{11}}}{W_{n}}& \dfrac{\widetilde{(T_{n})_{12}}}{W_{n}}\\ \\
\dfrac{\widetilde{(T_{n})_{21}}}{\widetilde{W_{n}}}& \dfrac{\widetilde{(T_{n})_{22}}}{\widetilde{W_{n}}}\\
\end{array} \right),
\end{equation}
with
\begin{equation}\label{fsst1}
W_{n}=\begin{vmatrix}
\lambda_{1}^{n}\phi_{1}&\lambda_{1}^{n-1}\varphi_{1}&\ldots&\lambda_{1}^{-(n-2)}\phi_{1}&\lambda_{1}^{-(n-1)}\varphi_{1}\\
\lambda_{2}^{n}\phi_{2}&\lambda_{2}^{n-1}\varphi_{2}&\ldots&\lambda_{2}^{-(n-2)}\phi_{2}&\lambda_{2}^{-(n-1)}\varphi_{2}\\
\vdots&\vdots&\vdots&\vdots&\vdots\\
\lambda_{2n-1}^{n}\phi_{2n-1}&\lambda_{2n-1}^{n-1}\varphi_{2n-1}&\ldots&\lambda_{2n-1}^{-(n-2)}\phi_{2n-1}&\lambda_{2n-1}^{-(n-1)}\varphi_{2n-1}\nonumber\\
\lambda_{2n}^{n}\phi_{2n}&\lambda_{2n}^{n-1}\varphi_{2n}&\ldots&\lambda_{2n}^{-(n-2)}\phi_{2n}&\lambda_{2n}^{-(n-1)}\varphi_{2n}\nonumber\\
\end{vmatrix},
\end{equation}
\begin{equation}\label{fsst2}
\widetilde{(T_{n})_{11}}=\begin{vmatrix}
\lambda^{n}&0&\ldots&\lambda^{-(n-2)}&0&\lambda^{-n}\\
\lambda_{1}^{n}\phi_{1}&\lambda_{1}^{n-1}\varphi_{1}&\ldots&\lambda_{1}^{-(n-2)}\phi_{1}&\lambda_{1}^{-(n-1)}\varphi_{1}&\lambda_{1}^{-n}\phi_{1}\\
\lambda_{2}^{n}\phi_{2}&\lambda_{2}^{n-1}\varphi_{2}&\ldots&\lambda_{2}^{-(n-2)}\phi_{2}&\lambda_{2}^{-(n-1)}\varphi_{2}&\lambda_{2}^{-n}\phi_{2}\\
\vdots&\vdots&\vdots&\vdots&\vdots&\vdots\\
\lambda_{2n-1}^{n}\phi_{2n-1}&\lambda_{2n-1}^{n-1}\varphi_{2n-1}&\ldots&\lambda_{2n-1}^{-(n-2)}\phi_{2n-1}&\lambda_{2n-1}^{-(n-1)}\varphi_{2n-1}&\lambda_{2n-1}^{-n}\phi_{2n-1}\nonumber\\
\lambda_{2n}^{n}\phi_{2n}&\lambda_{2n}^{n-1}\varphi_{2n}&\ldots&\lambda_{2n}^{-(n-2)}\phi_{2n}&\lambda_{2n}^{-(n-1)}\varphi_{2n}&\lambda_{2n}^{-n}\phi_{2n}\nonumber\\
\end{vmatrix},
\end{equation}
\begin{equation}\label{fsst3}
\widetilde{(T_{n})_{12}}=\begin{vmatrix}
0&\lambda^{n-1}&\ldots&0&\lambda^{-(n-1)}&0\\
\lambda_{1}^{n}\phi_{1}&\lambda_{1}^{n-1}\varphi_{1}&\ldots&\lambda_{1}^{-(n-2)}\phi_{1}&\lambda_{1}^{-(n-1)}\varphi_{1}&\lambda_{1}^{-n}\phi_{1}\\
\lambda_{2}^{n}\phi_{2}&\lambda_{2}^{n-1}\varphi_{2}&\ldots&\lambda_{2}^{-(n-2)}\phi_{2}&\lambda_{2}^{-(n-1)}\varphi_{2}&\lambda_{2}^{-n}\phi_{2}\\
\vdots&\vdots&\vdots&\vdots&\vdots&\vdots\\
\lambda_{2n-1}^{n}\phi_{2n-1}&\lambda_{2n-1}^{n-1}\varphi_{2n-1}&\ldots&\lambda_{2n-1}^{-(n-2)}\phi_{2n-1}&\lambda_{2n-1}^{-(n-1)}\varphi_{2n-1}&\lambda_{2n-1}^{-n}\phi_{2n-1}\nonumber\\
\lambda_{2n}^{n}\phi_{2n}&\lambda_{2n}^{n-1}\varphi_{2n}&\ldots&\lambda_{2n}^{-(n-2)}\phi_{2n}&\lambda_{2n}^{-(n-1)}\varphi_{2n}&\lambda_{2n}^{-n}\phi_{2n}\nonumber\\
\end{vmatrix},
\end{equation}
\begin{equation}\label{fsst4}
\widetilde{W_{n}}=\begin{vmatrix}
\lambda_{1}^{n}\varphi_{1}&\lambda_{1}^{n-1}\phi_{1}&\ldots&\lambda_{1}^{-(n-2)}\varphi_{1}&\lambda_{1}^{-(n-1)}\phi_{1}\\
\lambda_{2}^{n}\varphi_{2}&\lambda_{2}^{n-1}\phi_{2}&\ldots&\lambda_{2}^{-(n-2)}\varphi_{2}&\lambda_{2}^{-(n-1)}\phi_{2}\\
\vdots&\vdots&\vdots&\vdots&\vdots\\
\lambda_{2n-1}^{n}\varphi_{2n-1}&\lambda_{2n-1}^{n-1}\phi_{2n-1}&\ldots&\lambda_{2n-1}^{-(n-2)}\varphi_{2n-1}&\lambda_{2n-1}^{-(n-1)}\phi_{2n-1}\nonumber\\
\lambda_{2n}^{n}\varphi_{2n}&\lambda_{2n}^{n-1}\phi_{2n}&\ldots&\lambda_{2n}^{-(n-2)}\varphi_{2n}&\lambda_{2n}^{-(n-1)}\phi_{2n}\nonumber\\
\end{vmatrix},
\end{equation}
\begin{equation}\label{fsst5}
\widetilde{(T_{n})_{21}}=\begin{vmatrix}
0&\lambda^{n-1}&\ldots&0&\lambda^{-(n-1)}&0\\
\lambda_{1}^{n}\varphi_{1}&\lambda_{1}^{n-1}\phi_{1}&\ldots&\lambda_{1}^{-(n-2)}\varphi_{1}&\lambda_{1}^{-(n-1)}\phi_{1}&\lambda_{1}^{-n}\varphi_{1}\\
\lambda_{2}^{n}\varphi_{2}&\lambda_{2}^{n-1}\phi_{2}&\ldots&\lambda_{2}^{-(n-2)}\varphi_{2}&\lambda_{2}^{-(n-1)}\phi_{2}&\lambda_{2}^{-n}\varphi_{2}\\
\vdots&\vdots&\vdots&\vdots&\vdots&\vdots\\
\lambda_{2n-1}^{n}\varphi_{2n-1}&\lambda_{2n-1}^{n-1}\phi_{2n-1}&\ldots&\lambda_{2n-1}^{-(n-2)}\varphi_{2n-1}&\lambda_{2n-1}^{-(n-1)}\phi_{2n-1}&\lambda_{2n-1}^{-n}\varphi_{2n-1}\nonumber\\
\lambda_{2n}^{n}\varphi_{2n}&\lambda_{2n}^{n-1}\phi_{2n}&\ldots&\lambda_{2n}^{-(n-2)}\varphi_{2n}&\lambda_{2n}^{-(n-1)}\phi_{2n}&\lambda_{2n}^{-n}\varphi_{2n}\nonumber\\
\end{vmatrix},
\end{equation}
\begin{equation}\label{fsst6}
\widetilde{(T_{n})_{22}}=\begin{vmatrix}
\lambda^{n}&0&\ldots&\lambda^{-(n-2)}&0&\lambda^{-n}\\
\lambda_{1}^{n}\varphi_{1}&\lambda_{1}^{n-1}\phi_{1}&\ldots&\lambda_{1}^{-(n-2)}\varphi_{1}&\lambda_{1}^{-(n-1)}\phi_{1}&\lambda_{1}^{-n}\varphi_{1}\\
\lambda_{2}^{n}\varphi_{2}&\lambda_{2}^{n-1}\phi_{2}&\ldots&\lambda_{2}^{-(n-2)}\varphi_{2}&\lambda_{2}^{-(n-1)}\phi_{2}&\lambda_{2}^{-n}\varphi_{2}\\
\vdots&\vdots&\vdots&\vdots&\vdots&\vdots\\
\lambda_{2n-1}^{n}\varphi_{2n-1}&\lambda_{2n-1}^{n-1}\phi_{2n-1}&\ldots&\lambda_{2n-1}^{-(n-2)}\varphi_{2n-1}&\lambda_{2n-1}^{-(n-1)}\phi_{2n-1}&\lambda_{2n-1}^{-n}\varphi_{2n-1}\nonumber\\
\lambda_{2n}^{n}\varphi_{2n}&\lambda_{2n}^{n-1}\phi_{2n}&\ldots&\lambda_{2n}^{-(n-2)}\varphi_{2n}&\lambda_{2n}^{-(n-1)}\phi_{2n}&\lambda_{2n}^{-n}\varphi_{2n}\nonumber\\
\end{vmatrix},
\end{equation}

Next, we consider the transformed new solutions ($q^{[n]},r^{[n]}$) of FL system corresponding to
the n-fold DT. Under covariant requirement of spectral problem of the FL system, the transformed
form should be
\begin{equation}\label{sysn22}
\pa_{t}\psi^{[n]}=(J\lambda^2+Q^{[n]}\lambda+{V_{0}}^{[n]}+{V_{-1}}^{[n]}\lambda^{-1}+\dfrac{1}{4}J\lambda^{-2})\psi^{[n]}=V^{[n]}\psi^{[n]},
\end{equation}
with
\begin{equation}\label{fj1}
    \psi=\left( \begin{array}{c}
      \phi \\
      \varphi\\
     \end{array} \right),\nonumber\\
   \quad J= \left( \begin{array}{cc}
      -i &0 \\
      0 &i\\
   \end{array} \right),\nonumber\\
  \quad Q^{[n]}=\left( \begin{array}{cc}
     0 &{q_{x}}^{[n]} \\
     {r_{x}}^{[n]} &0\\
  \end{array} \right),\nonumber\\
\end{equation}
\begin{equation}\label{fj2}
{V_{0}}^{[n]}=\left( \begin{array}{cc}
i-\frac{1}{2}iq^{[n]}r^{[n]} &0 \\
 0&-i+\frac{1}{2}iq^{[n]}r^{[n]}\\
\end{array} \right),\quad {V_{-1}}^{[n]}=\left( \begin{array}{cc}
0 &\frac{1}{2}iq^{[n]} \\
- \frac{1}{2}ir^{[n]}&0\\
\end{array} \right).\nonumber\\
\end{equation}
and then  \begin{equation}\label{ntt2} {T_{n}}_{t}+T_{n}~V=V^{[n]}~T_{n}.
\end{equation}
Substituting $T_n$ given by eq.(\ref{tnss}) into eq.(\ref{ntt2}), and then comparing the
coefficients of $\lambda^{-(n+1)}$, it yields
\begin{eqnarray}\label{ntt3}
&&q^{[1]}=q +b_{-(n-1)},
\ \ r^{[1]}=r+c_{-(n-1)}.
\end{eqnarray}
Furthermore, taking $b_{-(n-1)},c_{-(n-1)}$
which are obtained from eq.(\ref{fss1}), then new solutions ($q^{[n]},r^{[n]}$)
are given by
\begin{eqnarray}\label{ntt4}
&&q^{[n]}=q+\dfrac{\Omega_{-(n-1)}}{W_{n}}, \ \
r^{[n]}=r+\dfrac{\widetilde{\Omega}_{-(n-1)}}{\widetilde{W_{n}}}.
\end{eqnarray}
with
\begin{equation*}
\Omega_{-(n-1)}=\begin{vmatrix}
\lambda_{1}^{n}\phi_{1}&\lambda_{1}^{n-1}\varphi_{1}&\ldots&\lambda_{1}^{-(n-2)}\phi_{1}&-\lambda_{1}^{-n}\phi_{1}\\
\lambda_{2}^{n}\phi_{2}&\lambda_{2}^{n-1}\varphi_{2}&\ldots&\lambda_{2}^{-(n-2)}\phi_{2}&-\lambda_{2}^{-n}\phi_{2}\\
\vdots&\vdots&\vdots&\vdots&\vdots\\
\lambda_{2n-1}^{n}\phi_{2n-1}&\lambda_{2n-1}^{n-1}\varphi_{2n-1}&\ldots&\lambda_{2n-1}^{-(n-2)}\phi_{2n-1}&-\lambda_{2n-1}^{-n}\phi_{2n-1}\nonumber\\
\lambda_{2n}^{n}\phi_{2n}&\lambda_{2n}^{n-1}\varphi_{2n}&\ldots&\lambda_{2n}^{-(n-2)}\phi_{2n}&-\lambda_{2n}^{-n}\phi_{2n}\nonumber\\
\end{vmatrix}
\end{equation*}
\begin{equation*}
\widetilde{\Omega}_{-(n-1)}=\begin{vmatrix}
\lambda_{1}^{n}\varphi_{1}&\lambda_{1}^{n-1}\phi_{1}&\ldots&\lambda_{1}^{-(n-2)}\varphi_{1}&-\lambda_{1}^{-n}\varphi_{1}\\
\lambda_{2}^{n}\varphi_{2}&\lambda_{2}^{n-1}\phi_{2}&\ldots&\lambda_{2}^{-(n-2)}\varphi_{2}&-\lambda_{2}^{-n}\varphi_{2}\\
\vdots&\vdots&\vdots&\vdots&\vdots\\
\lambda_{2n-1}^{n}\varphi_{2n-1}&\lambda_{2n-1}^{n-1}\phi_{2n-1}&\ldots&\lambda_{2n-1}^{-(n-2)}\varphi_{2n-1}&-\lambda_{2n-1}^{-n}\varphi_{2n-1}\nonumber\\
\lambda_{2n}^{n}\varphi_{2n}&\lambda_{2n}^{n-1}\phi_{2n}&\ldots&\lambda_{2n}^{-(n-2)}\varphi_{2n}&-\lambda_{2n}^{-n}\varphi_{2n}\nonumber\\
\end{vmatrix}.
\end{equation*}

and the new eigenfunction $\psi_k^{[n]}$ of $\lambda_k$
is\begin{equation}
\psi^{[n]}_k\mbox{\hspace{-0.15cm}}=\mbox{\hspace{-0.15cm}} \left(
\begin{array}{c}\mbox{\hspace{-0.15cm}}
\dfrac{
\begin{vmatrix}
\lambda_{k}^{n}\phi_{k}&\lambda_{k}^{n-1}\varphi_{k}&\ldots&\lambda_{k}^{-(n-2)}\phi_{k}&\lambda_{k}^{-(n-1)}\varphi_{k}&\lambda_{k}^{-n}\phi_{k}\\
\lambda_{1}^{n}\phi_{1}&\lambda_{1}^{n-1}\varphi_{1}&\ldots&\lambda_{1}^{-(n-2)}\phi_{1}&\lambda_{1}^{-(n-1)}\varphi_{1}&\lambda_{1}^{-n}\phi_{1}\\
\lambda_{2}^{n}\phi_{2}&\lambda_{2}^{n-1}\varphi_{2}&\ldots&\lambda_{2}^{-(n-2)}\phi_{2}&\lambda_{2}^{-(n-1)}\varphi_{2}&\lambda_{2}^{-n}\phi_{2}\\
\vdots&\vdots&\vdots&\vdots&\vdots&\vdots\\
\lambda_{2n-1}^{n}\phi_{2n-1}&\lambda_{2n-1}^{n-1}\varphi_{2n-1}&\ldots&\lambda_{2n-1}^{-(n-2)}\phi_{2n-1}&\lambda_{2n-1}^{-(n-1)}\varphi_{2n-1}&\lambda_{2n-1}^{-n}\phi_{2n-1}\nonumber\\
\lambda_{2n}^{n}\phi_{2n}&\lambda_{2n}^{n-1}\varphi_{2n}&\ldots&\lambda_{2n}^{-(n-2)}\phi_{2n}&\lambda_{2n}^{-(n-1)}\varphi_{2n}&\lambda_{2n}^{-n}\phi_{2n}\nonumber\\
\end{vmatrix}}{|W_{n}|}\\
\mbox{\hspace{-0.15cm}}\dfrac{\begin{vmatrix}
\lambda_{k}^{n}\varphi_{k}&\lambda_{k}^{n-1}\phi_{k}&\ldots&\lambda_{k}^{-(n-2)}\varphi_{k}&\lambda_{k}^{-(n-1)}\phi_{k}&\lambda_{k}^{-n}\varphi_{k}\\
\lambda_{1}^{n}\varphi_{1}&\lambda_{1}^{n-1}\phi_{1}&\ldots&\lambda_{1}^{-(n-2)}\varphi_{1}&\lambda_{1}^{-(n-1)}\phi_{1}&\lambda_{1}^{-n}\varphi_{1}\\
\lambda_{2}^{n}\varphi_{2}&\lambda_{2}^{n-1}\phi_{2}&\ldots&\lambda_{2}^{-(n-2)}\varphi_{2}&\lambda_{2}^{-(n-1)}\phi_{2}&\lambda_{2}^{-n}\varphi_{2}\\
\vdots&\vdots&\vdots&\vdots&\vdots&\vdots\\
\lambda_{2n-1}^{n}\varphi_{2n-1}&\lambda_{2n-1}^{n-1}\phi_{2n-1}&\ldots&\lambda_{2n-1}^{-(n-2)}\varphi_{2n-1}&\lambda_{2n-1}^{-(n-1)}\phi_{2n-1}&\lambda_{2n-1}^{-n}\varphi_{2n-1}\nonumber\\
\lambda_{2n}^{n}\varphi_{2n}&\lambda_{2n}^{n-1}\phi_{2n}&\ldots&\lambda_{2n}^{-(n-2)}\varphi_{2n}&\lambda_{2n}^{-(n-1)}\phi_{2n}&\lambda_{2n}^{-n}\varphi_{2n}\nonumber\\
\end{vmatrix}}{|\widetilde{W_{n}}|}
\end{array}.
\right)
\end{equation}

We are now in a position to consider the reduction of the DT of the FL system so that
$q^{[n]}=(r^{[n]})^*$, then the DT of the FL equation is given. Under the reduction condition $q=r^*$,
the eigenfunction $\psi_k=\left( \begin{array}{c}
\phi_k\\
\varphi_k
\end{array} \right)$ associated with eigenvalue $\lambda_k$ has following properties,
${\phi_{k}}^{\ast}=\varphi_{l}, \ {\varphi_{k}}^{\ast}=\phi_{l}$,
${\lambda_{k}}^{\ast}=\lambda_{l}$, where $k\neq l$. For example,
setting $l=1,3,\dots,2n-1$, then choosing the $n$ distinct
eigenvalues and eigenfunctions in n-fold DTs in the following manner:
\begin{equation}\label{2nfoldredu}
 \lambda_l \leftrightarrow \psi_l=\left( \begin{array}{c}
\phi_l\\
\varphi_l\end{array} \right)
, \text{and} \lambda_{2l}= \lambda_{2l-1}^*,\leftrightarrow
\psi_{2l}=\left( \begin{array}{c}
\varphi_{2l-1}^*\\
\phi_{2l-1}^*
\end{array} \right)
\end{equation}
so that $q^{[n]}=(r^{[n]})^*$ in eq.(\ref{ntt4}). Then $T_{n}$ with
these paired-eigenvalues $\lambda_i$ and paired-eigenfunctions
$\psi_i(i=1,3,\dots,2n-1)$ is reduced to the n-fold DT of the FL
equation. Notice that the denominator $W_n$ of $q^{[n]}$ is a
modulus of a non-zero complex function under reduction condition, so
the new solution $q^{[n]}$ is non-singular.

\section{The n-order rogue waves and their determinant forms}
Using the results of DT above, breather solutions of FL equation are
generated by assuming a periodic seed solution, then we can
construct the  rogue waves of the FL equation from a Taylor series
expansion of  the breather solutions.

Set $a$ and  $c$ be two complex constants,
then $ q=c \exp{(i(a x+(\dfrac{(a+1)^2}{a}-c^2)t))}$ is a periodic solution of the FL equation, which will be used as
a seed solution of the DT. Substituting $q=c \exp{(i(a x+(\dfrac{(a+1)^2}{a}-c^2)t))}$ into the spectral problem eq.(\ref{sys11})
 and eq.(\ref{sys22}), and using the method of separation of variables and the superposition
principle, the eigenfunction $\psi_{2k-1}$ associated with
$\lambda_{2k-1}$ is given by
\begin{eqnarray}\label{eigenfunfornonzeroseed}
\left(\mbox{\hspace{-0.2cm}} \begin{array}{c}
 \mbox{\hspace{-0.1cm}}\phi_{2k-1}(x,t,\lambda_{2k-1})\mbox{\hspace{-0.1cm}}\\
 \mbox{\hspace{-0.1cm}}\varphi_{2k-1}(x,t,\lambda_{2k-1})\mbox{\hspace{-0.1cm}}\\
\end{array}\mbox{\hspace{-0.2cm}}\right)\mbox{\hspace{-0.25cm}}=\mbox{\hspace{-0.25cm}}\left(\mbox{\hspace{-0.2cm}}\begin{array}{c}
 \mbox{\hspace{-0.1cm}}C_1\varpi_1(x,t,\lambda_{2k-1})[1]\mbox{\hspace{-0.15cm}}+\mbox{\hspace{-0.15cm}}C_2\varpi_2(x,t,\lambda_{2k-1})[1]\mbox{\hspace{-0.15cm}}+\mbox{\hspace{-0.15cm}}C_3\varpi_1^{\ast}(x,t,{\lambda_{2k-1}^{\ast})}[2]\mbox{\hspace{-0.15cm}}+\mbox{\hspace{-0.15cm}}C_4\varpi_2^{\ast}(x,t,{\lambda_{2k-1}^{\ast})}[2]\mbox{\hspace{-0.1cm}}\\
 \mbox{\hspace{-0.1cm}}C_1\varpi_1(x,t,\lambda_{2k-1})[2]\mbox{\hspace{-0.15cm}}+\mbox{\hspace{-0.15cm}}C_2\varpi_2(x,t,\lambda_{2k-1})[2]\mbox{\hspace{-0.15cm}}+\mbox{\hspace{-0.15cm}}C_3\varpi_1^{\ast}(x,t,{\lambda_{2k-1}^{\ast})}[1]\mbox{\hspace{-0.15cm}}+\mbox{\hspace{-0.15cm}}C_4\varpi_2^{\ast}(x,t,{\lambda_{2k-1}^{\ast})}[1]\mbox{\hspace{-0.1cm}}\\
\end{array}\mbox{\hspace{-0.2cm}}\right).
\end{eqnarray}
Here
\begin{eqnarray*}
\left(\mbox{\hspace{-0.2cm}}\begin{array}{c}
 \varpi_1(x,t,\lambda_{2k-1})[1]\\
 \varpi_1(x,t,\lambda_{2k-1})[2]\\
\end{array}\mbox{\hspace{-0.2cm}}\right)\mbox{\hspace{-0.25cm}}=\mbox{\hspace{-0.25cm}}\left(\mbox{\hspace{-0.2cm}}\begin{array}{c}
\exp(\sqrt{S(\lambda_{2k-1})}( \dfrac{1}{2}x\mbox{\hspace{-0.15cm}}+\mbox{\hspace{-0.15cm}}\dfrac{2a{\lambda_{2k-1}}^{2}\mbox{\hspace{-0.15cm}}+\mbox{\hspace{-0.15cm}}1}{4a{\lambda_{2k-1}}^{2}} t)\mbox{\hspace{-0.15cm}}+\mbox{\hspace{-0.15cm}}\dfrac{1}{2}i(a x\mbox{\hspace{-0.15cm}}+\mbox{\hspace{-0.15cm}}(\dfrac{(a\mbox{\hspace{-0.15cm}}+\mbox{\hspace{-0.15cm}}1)^2}{a}\mbox{\hspace{-0.15cm}}-\mbox{\hspace{-0.15cm}}c^2)t)) \\
\dfrac{ a +2 {\lambda_{2k-1}}^{2}-i\sqrt{S(\lambda_{2k-1})}}{2{\lambda_{2k-1}} ca}\exp(\sqrt{S(\lambda_{2k-1})}( \dfrac{1}{2}x\mbox{\hspace{-0.15cm}}+\mbox{\hspace{-0.15cm}}\dfrac{2a{\lambda_{2k-1}}^{2}\mbox{\hspace{-0.15cm}}+\mbox{\hspace{-0.15cm}}1}{4a{\lambda_{2k-1}}^{2}} t)\mbox{\hspace{-0.15cm}}-\mbox{\hspace{-0.15cm}}\dfrac{1}{2}i(a x\mbox{\hspace{-0.15cm}}+\mbox{\hspace{-0.15cm}}(\dfrac{(a\mbox{\hspace{-0.15cm}}+\mbox{\hspace{-0.15cm}}1)^2}{a}\mbox{\hspace{-0.15cm}}-\mbox{\hspace{-0.15cm}}c^2)t)) \\
\end{array}\mbox{\hspace{-0.2cm}}\right),\\
\end{eqnarray*}
\begin{eqnarray*}
\left(\mbox{\hspace{-0.2cm}}\begin{array}{c}
 \varpi_2(x,t,\lambda_{2k-1})[1]\\
 \varpi_2(x,t,\lambda_{2k-1})[2]\\
\end{array}\mbox{\hspace{-0.2cm}}\right)\mbox{\hspace{-0.25cm}}=\mbox{\hspace{-0.25cm}}\left(\mbox{\hspace{-0.2cm}}\begin{array}{c}
\exp(-\sqrt{S(\lambda_{2k-1})}( \dfrac{1}{2}x\mbox{\hspace{-0.15cm}}+\mbox{\hspace{-0.15cm}}\dfrac{2a{\lambda_{2k-1}}^{2}\mbox{\hspace{-0.15cm}}+\mbox{\hspace{-0.15cm}}1}{4a{\lambda_{2k-1}}^{2}} t)\mbox{\hspace{-0.15cm}}+\mbox{\hspace{-0.15cm}}\dfrac{1}{2}i(a x\mbox{\hspace{-0.15cm}}+\mbox{\hspace{-0.15cm}}(\dfrac{(a\mbox{\hspace{-0.15cm}}+\mbox{\hspace{-0.15cm}}1)^2}{a}\mbox{\hspace{-0.15cm}}-\mbox{\hspace{-0.15cm}}c^2)t)) \\
\dfrac{ a +2 {\lambda_{2k-1}}^{2}+i\sqrt{S(\lambda_{2k-1})}}{2{\lambda_{2k-1}} ca}\exp(-\sqrt{S(\lambda_{2k-1})}( \dfrac{1}{2}x\mbox{\hspace{-0.15cm}}+\mbox{\hspace{-0.15cm}}\dfrac{2a{\lambda_{2k-1}}^{2}\mbox{\hspace{-0.15cm}}+\mbox{\hspace{-0.15cm}}1}{4a{\lambda_{2k-1}}^{2}} t)\mbox{\hspace{-0.15cm}}-\mbox{\hspace{-0.15cm}}\dfrac{1}{2}i(a x\mbox{\hspace{-0.15cm}}+\mbox{\hspace{-0.15cm}}(\dfrac{(a\mbox{\hspace{-0.15cm}}+\mbox{\hspace{-0.15cm}}1)^2}{a}\mbox{\hspace{-0.15cm}}-\mbox{\hspace{-0.15cm}}c^2)t)) \\
\end{array}\mbox{\hspace{-0.2cm}}\right),\\
\end{eqnarray*}
\begin{eqnarray*}
\varpi_1(x,t,\lambda_{2k-1})= \left( \begin{array}{c}
 \varpi_1(x,t,\lambda_{2k-1})[1]\\
 \varpi_1(x,t,\lambda_{2k-1})[2]\\
\end{array} \right),~~~~~
\varpi_2(x,t,\lambda_{2k-1})= \left( \begin{array}{c}
 \varpi_2(x,t,\lambda_{2k-1})[1]\\
 \varpi_2(x,t,\lambda_{2k-1})[2]\\
\end{array} \right),
\end{eqnarray*}
\begin{eqnarray}
&S(\lambda_{2k-1})=-a^{2}-4{\lambda_{2k-1}}^{4}+4{\lambda_{2k-1}}^{2}(c^{2}a^{2}-a)
(k=1,2,\dots,n). \nonumber
\end{eqnarray}
Here $a,c,x,t \in \Bbb R$, $C_1, C_2, C_3, C_4 \in \Bbb C$. Note
that $\varpi_1(x,t,\lambda_{2k-1})$ and
$\varpi_2(x,t,\lambda_{2k-1})$ are two different solutions of the
spectral problem eq.(\ref{sys11}) and eq.(\ref{sys22}), but we  can
only get the  trivial solutions through DT of the FL equation by
setting eigenfunction $\psi_{2k-1}$ be one of them.
What is more, we can get  richer solutions by using (\ref{eigenfunfornonzeroseed}).\\

3.1 The first-order rogue waves generated by first-order breather
solutions\\

 Under the choice in eq.(\ref{2nfoldredu}) with one paired eigenvalue
$\lambda_{1}=\alpha_{1}+i\beta_{1}$ and $\lambda_{2}=\alpha_{1}-i\beta_{1}$,
the two-fold DT eq.(\ref{ntt4}) of the FL equation implies  a solution
\begin{equation}\label{q2j2}
 q^{[1]}=q+\dfrac{({\lambda_{2}}^{2}-{\lambda_{1}}^{2})\phi_{1}{\varphi_{1}}^{\ast}}{\lambda_2\lambda_1(-\lambda_{2}\varphi_{1}{\varphi_{1}}^{\ast}+\lambda_{1}\phi_{1}{\phi_{1}}^{\ast})},
\end{equation}
with $\phi_{1}$ and $\varphi_{1}$ given by eq.(\ref{eigenfunfornonzeroseed}).
For simplicity, under the condition $C_1=C_2=C_3=C_4=1$, let
$c=-\dfrac{\sqrt{a+2{\alpha_1}^{2}-2{\beta_1}^{2}}}{a}$ so that
$\text{Im}(-a^{2}-4{\lambda_{2k-1}}^{4}+4{\lambda_{2k-1}}^{2}
(c^{2}a^{2}-a))=0$, we get the first order breather
$q^{[1]}$\cite{jhswk}. Furthermore, by letting $ a \rightarrow
2({\alpha_{1}}^{2}+{\beta_{1}}^{2})$ in
 $(\ref{q2j2})$ with $\text{Im}(-a^{2}-4{\lambda_{2k-1}}^{4}+4{\lambda_{2k-1}}^{2}(c^{2}a^{2}-a))=0$,
its first order breather $q^{[1]}$ becomes a rogue wave
$q^{[1]}_{rw}$\cite{jhswk}.\\

3.2 The n-order rogue waves and their determinant forms\\

In order to make the higher order rogue waves informative, we modify
 $C_1, C_2, C_3$ and $C_4$ in the equation (\ref{eigenfunfornonzeroseed})
\begin{eqnarray}
&&C_1=K_0+\exp(-\dfrac{1}{2}iS(\lambda_{2k-1})\sum_{j=0}^{k-1}J_j(\lambda_{2k-1}-\lambda_0)^{j})\nonumber\\
&&C_2=K_0+\exp(\dfrac{1}{2}iS(\lambda_{2k-1})\sum_{j=0}^{k-1}J_j(\lambda_{2k-1}-\lambda_0)^{j})\nonumber\\
&&C_3=K_0+\exp(-\dfrac{1}{2}iS(\lambda_{2k-1})\sum_{j=0}^{k-1}L_j(\lambda_{2k-1}-\lambda_0)^{j})\nonumber\\
&&C_4=K_0+\exp(\dfrac{1}{2}iS(\lambda_{2k-1})\sum_{j=0}^{k-1}L_j(\lambda_{2k-1}-\lambda_0)^{j})\label{chooseCC}
\end{eqnarray}
Here $K_0,J_j,L_j \in \Bbb C$. Note that
$\lambda_{2k-1}=\lambda_0=-\dfrac{1}{2}a c +i\dfrac{1}{2}\sqrt{-a^2
c^2+2a}$ is the zero point of $S(\lambda_{2k-1})$.

In this way, higher order rogue waves can be
constructed from the higher breather solutions. In other words, let $
\lambda_{2k-1}\rightarrow -\dfrac{1}{2}a c +i\dfrac{1}{2}\sqrt{-a^2
c^2+2a}$ in n-order breather solutions, n-order rogue waves can be
given. Generally, in comparison to the method of limiting the breather
solutions, the method of making rational eigenfunction below may be
more direct and the rogue wave can be shown in determinant forms.

Substituting eq.(\ref{chooseCC}) into
eqs.(\ref{eigenfunfornonzeroseed}), by assuming $
\lambda_{2k-1}\rightarrow -\dfrac{1}{2}a c +i\dfrac{1}{2}\sqrt{-a^2
c^2+2a}$,  eigenfunction $\psi_{2k-1}$ associated with
$\lambda_{2k-1}$ become rational eigenfunction $\psi_r$. For
simplicity, when $ a=1, c=-1$, rational eigenfunction $\psi_r$ has
the following form

\begin{eqnarray*}
\left(\mbox{\hspace{-0.2cm}} \begin{array}{c}
 \phi_{r}\\
 \varphi_{r}\\
\end{array}\mbox{\hspace{-0.2cm}}\right)\mbox{\hspace{-0.2cm}}=\mbox{\hspace{-0.2cm}}\left(\mbox{\hspace{-0.2cm}}\begin{array}{c}
 \mbox{\hspace{-0.1cm}}((-2-2i)K x-4Kt+(-1+i)(-L_0
^2+J_0^2+2i+2iK_0-2iL_0+2L_0))\exp(\dfrac{1}{2}i(x+3t))\\
 \mbox{\hspace{-0.1cm}}((2+2i)K x+4K t+(1-i)(-2i-
L_0^2+J_0^2-2iK_0-2iJ_0+2J_0))\exp(-\dfrac{1}{2}i(x+3t))\\
\end{array}\mbox{\hspace{-0.2cm}}\right),
\end{eqnarray*}
\begin{eqnarray}\label{reigenfunfornonzeroseed}
K=-L_0+1-i-iK_0+K_0+J_0
\end{eqnarray}

Substituting eigenfunctions eq.(\ref{reigenfunfornonzeroseed}) into
eqs.(\ref{ntt4}), we can get the first order rogue wave
${q}^{[1]}_{rw}$ in the form of determinant.  The dynamical
evolution of ${q}^{[1]}_{rw}$ for the parametric choice $K_0=1,
J_0=L_0=10 $ is plotted in the Figure 1, which control the position
of the first-order rogue wave by choosing the parameters $K_0, J_0$
and $L_0$. Similarly, the corresponding density plot is portrayed in
the Figure 2. \noindent {\bf Theorem 3.} For the n-fold DT, the
n-order rogue wave ${q}^{[n]}_{rw}$ is of the form
\begin{eqnarray}\label{rntt4}
&&q_{rw}^{[n]}=q+\dfrac{\Omega_{r-(n-1)}}{W_{rn}}.
\end{eqnarray}
with
\begin{equation}\label{ntt5}
W_{rn}=\begin{vmatrix}
h_{n1}^1&h_{n-12}^1&h_{n-21}^1&h_{n-32}^1&\ldots&h_{-(n-2)1}^1&h_{-(n-1)2}^1\\
{h_{n2}^1}^{*}&{h_{n-11}^1}^{*}&{h_{n-22}^1}^{*}&{h_{n-31}^1}^{*}&\ldots&{h_{-(n-2)2}^1}^{*}&{h_{-(n-1)1}^1}^{*}\\
\vdots&\vdots&\vdots&\vdots&\vdots&\vdots&\vdots\\
h_{n1}^n&h_{n-12}^n&h_{n-21}^n&h_{n-32}^n&\ldots&h_{-(n-2)1}^n&h_{-(n-1)2}^n\\
{h_{n2}^n}^{*}&{h_{n-11}^n}^{*}&{h_{n-22}^n}^{*}&{h_{n-31}^n}^{*}&\ldots&{h_{-(n-2)2}^n}^{*}&{h_{-(n-1)1}^n}^{*}\nonumber\\
\end{vmatrix},
\end{equation}
\begin{equation*}
\Omega_{r-(n-1)}=\begin{vmatrix}
h_{n1}^1&h_{n-12}^1&h_{n-21}^1&h_{n-32}^1&\ldots&h_{-(n-2)1}^1&-h_{-n1}^1\\
{h_{n2}^1}^{*}&{h_{n-11}^1}^{*}&{h_{n-22}^1}^{*}&{h_{n-31}^1}^{*}&\ldots&{h_{-(n-2)2}^1}^{*}&-{h_{-n2}^1}^{*}\\
\vdots&\vdots&\vdots&\vdots&\vdots&\vdots&\vdots\\
h_{n1}^n&h_{n-12}^n&h_{n-21}^n&h_{n-32}^n&\ldots&h_{-(n-2)1}^n&-h_{-n1}^n\\
{h_{n2}^n}^{*}&{h_{n-11}^n}^{*}&{h_{n-22}^n}^{*}&{h_{n-31}^n}^{*}&\ldots&{h_{-(n-2)2}^n}^{*}&-{h_{-n2}^n}^{*}\nonumber\\
\end{vmatrix},
\end{equation*}
The final form of $\bar{T}_{rn}(\lambda)$ has the
form,\begin{equation}
\label{rfss1}T_{rn}=T_{rn}(\lambda;\lambda_1,\lambda_2,\cdots,\lambda_{2n})=\left(
\begin{array}{cc}
\dfrac{\widetilde{(T_{rn})_{11}}}{W_{rn}}& \dfrac{\widetilde{(T_{rn})_{12}}}{W_{rn}}\\ \\
\dfrac{\widetilde{(T_{rn})_{21}}}{\widetilde{W_{rn}}}& \dfrac{\widetilde{(T_{rn})_{22}}}{\widetilde{W_{rn}}}\\
\end{array} \right),
\end{equation}
with
\begin{equation}\label{fsst2}
\widetilde{(T_{rn})_{11}}=\begin{vmatrix}
\lambda^{n}&0&\lambda^{n-2}&0&\ldots&\lambda^{-(n-2)}&0&\lambda^{-n}\\
h_{n1}^1&h_{n-12}^1&h_{n-21}^1&h_{n-32}^1&\ldots&h_{-(n-2)1}^1&h_{-(n-1)2}^1&h_{-n1}^1\\
{h_{n2}^1}^{*}&{h_{n-11}^1}^{*}&{h_{n-22}^1}^{*}&{h_{n-31}^1}^{*}&\ldots&{h_{-(n-2)2}^1}^{*}&{h_{-(n-1)1}^1}^{*}&{h_{-n2}^1}^{*}\\
\vdots&\vdots&\vdots&\vdots&\vdots&\vdots&\vdots&\vdots\\
h_{n1}^n&h_{n-12}^n&h_{n-21}^n&h_{n-32}^n&\ldots&h_{-(n-2)1}^n&h_{-(n-1)2}^n&h_{-n1}^n\\
{h_{n2}^n}^{*}&{h_{n-11}^n}^{*}&{h_{n-22}^n}^{*}&{h_{n-31}^n}^{*}&\ldots&{h_{-(n-2)2}^n}^{*}&{h_{-(n-1)1}^n}^{*}&{h_{-n2}^n}^{*}\nonumber\\
\end{vmatrix},
\end{equation}
\begin{equation}\label{fsst3}
\widetilde{(T_{rn})_{12}}=\begin{vmatrix}
0&\lambda^{n-1}&0&\lambda^{n-3}&\ldots&0&\lambda^{-(n-1)}&0\\
h_{n1}^1&h_{n-12}^1&h_{n-21}^1&h_{n-32}^1&\ldots&h_{-(n-2)1}^1&h_{-(n-1)2}^1&h_{-n1}^1\\
{h_{n2}^1}^{*}&{h_{n-11}^1}^{*}&{h_{n-22}^1}^{*}&{h_{n-31}^1}^{*}&\ldots&{h_{-(n-2)2}^1}^{*}&{h_{-(n-1)1}^1}^{*}&{h_{-n2}^1}^{*}\\
\vdots&\vdots&\vdots&\vdots&\vdots&\vdots&\vdots&\vdots\\
h_{n1}^n&h_{n-12}^n&h_{n-21}^n&h_{n-32}^n&\ldots&h_{-(n-2)1}^n&h_{-(n-1)2}^n&h_{-n1}^n\\
{h_{n2}^n}^{*}&{h_{n-11}^n}^{*}&{h_{n-22}^n}^{*}&{h_{n-31}^n}^{*}&\ldots&{h_{-(n-2)2}^n}^{*}&{h_{-(n-1)1}^n}^{*}&{h_{-n2}^n}^{*}\nonumber\\
\end{vmatrix},
\end{equation}
\begin{equation}\label{fsst5}
\widetilde{(T_{rn})_{21}}=\begin{vmatrix}
0&\lambda^{n-1}&0&\lambda^{n-3}&\ldots&0&\lambda^{-(n-1)}&0\\
h_{n2}^1&h_{n-11}^1&h_{n-22}^1&h_{n-31}^1&\ldots&h_{-(n-2)2}^1&h_{-(n-1)1}^1&h_{-n2}^1\\
{h_{n1}^1}^{*}&{h_{n-12}^1}^{*}&{h_{n-21}^1}^{*}&{h_{n-32}^1}^{*}&\ldots&{h_{-(n-2)1}^1}^{*}&{h_{-(n-1)2}^1}^{*}&{h_{-n1}^1}^{*}\\
\vdots&\vdots&\vdots&\vdots&\vdots&\vdots&\vdots&\vdots\\
h_{n2}^n&h_{n-11}^n&h_{n-22}^n&h_{n-31}^n&\ldots&h_{-(n-2)2}^n&h_{-(n-1)1}^n&h_{-n2}^n\\
{h_{n1}^n}^{*}&{h_{n-12}^n}^{*}&{h_{n-21}^n}^{*}&{h_{n-32}^n}^{*}&\ldots&{h_{-(n-2)1}^n}^{*}&{h_{-(n-1)2}^n}^{*}&{h_{-n1}^n}^{*}\nonumber\\
\end{vmatrix},
\end{equation}
\begin{equation}\label{fsst6}
\widetilde{(T_{rn})_{22}}=\begin{vmatrix}
\lambda^{n}&0&\lambda^{n-2}&0&\ldots&\lambda^{-(n-2)}&0&\lambda^{-n}\\
h_{n2}^1&h_{n-11}^1&h_{n-22}^1&h_{n-31}^1&\ldots&h_{-(n-2)2}^1&h_{-(n-1)1}^1&h_{-n2}^1\\
{h_{n1}^1}^{*}&{h_{n-12}^1}^{*}&{h_{n-21}^1}^{*}&{h_{n-32}^1}^{*}&\ldots&{h_{-(n-2)1}^1}^{*}&{h_{-(n-1)2}^1}^{*}&{h_{-n1}^1}^{*}\\
\vdots&\vdots&\vdots&\vdots&\vdots&\vdots&\vdots&\vdots\\
h_{n2}^n&h_{n-11}^n&h_{n-22}^n&h_{n-31}^n&\ldots&h_{-(n-2)2}^n&h_{-(n-1)1}^n&h_{-n2}^n\\
{h_{n1}^n}^{*}&{h_{n-12}^n}^{*}&{h_{n-21}^n}^{*}&{h_{n-32}^n}^{*}&\ldots&{h_{-(n-2)1}^n}^{*}&{h_{-(n-1)2}^n}^{*}&{h_{-n1}^n}^{*}\nonumber\\
\end{vmatrix},
\end{equation}
Here
\begin{eqnarray*}
h_{m1}^{l}=\dfrac{\partial^{l}}{\partial{\delta}^{l}}((\lambda_0+\delta)^{m}\phi_1(\lambda_1=\lambda_0+\delta))|_{\delta=0},
m=-n,-(n-1),\cdots,-1,0,1,\cdots,n-1, n,l=1,2,\cdots 2n.
\end{eqnarray*}
\begin{eqnarray*}
h_{m2}^{l}=\dfrac{\partial^{l}}{\partial{\delta}^{l}}((\lambda_0+\delta)^{m}\varphi_1(\lambda_1=\lambda_0+\delta))|_{\delta=0},
m=-n,-(n-1),\cdots,-1,0,1,\cdots,n-1, n,l=1,2,\cdots 2n.
\end{eqnarray*}

\noindent {\bf Case 1).}  When $n=2$, substituting
eq.(\ref{eigenfunfornonzeroseed}) into eq.(\ref{rntt4}),
 one can construct the
second-order rogue waves with seven free parameters.  Note that
under the condition $J_1>>J_0$ and $L_1>>L_0$ except for $J_0=-L_0,
J_1=-L_1$, the second-rogue can split into three first-order rogue
wave (triplets rogue wave) \cite{triplets} rather than two. The
dynamical evolution of $|{q}^{[2]}_{rw}|^{2}$ for the parametric
choice $K_0=1, J_0=-L_0, J_1=L_1, L_0=-5, L_1=20, a=1, c=-1$ is
plotted in the Figure. 3 and the corresponding density plot is shown
in the Figure. 4. There is another kind second-order rogue wave, for
example, $|{q}^{[2]}_{rw}|^{2}$ is higher than second-rogue above,
which is a a fundamental pattern\cite{hefokas1}. The dynamical
evolution of $|{q}^{[2]}_{rw}|^{2}$ for the parametric choice
$J_0=-L_0, J_1=-L_1, a=1, c=-1$ is plotted in the Figure. 5 and the
corresponding density map is portrayed in Figure. 6. Note that
$|{q}^{[2]}_{rw}|^{2}$ has only two free parameters $a$ and $c$
under the condition $J_0=-L_0, J_1=-L_1$.

\noindent {\bf Case 2).}  When $n=3$, substituting
eq.(\ref{eigenfunfornonzeroseed}) into eq.(\ref{rntt4}) can lead to the
third-order rogue waves with nine free parameters. Note that under
the condition $J_2>>J_i, L_2>>L_i (i=0,1)$  or $J_1>>J_i, L_1>>L_i
 (i=0,2)$ except for $J_0=-L_0, J_1=-L_1,
J_2=-L_2$, the third-rogue can split into six first-order rogue wave
rather. Circular rogue wave \cite{Circular} may be constructed by
the condition $J_2>>J_i, L_2>>L_i (i=0,1)$ except for $J_0=-L_0,
J_1=-L_1, J_2=-L_2$. The dynamical evolution of
$|{q}^{[3]}_{rw}|^{2}$(circular pattern) for the parametric choice
$K_0=1, J_0=-L_0, J_1=-L_1, J_2=L_2, L_0=5, L_1=5, L_2=3000, a=1,
c=-1$ is plotted in the Figure. 7 and the corresponding density plot
in Figure. 8. At the same time, triplets rogue wave  may be
constructed by the condition $J_1>>J_i, L_1>>L_i (i=0,2)$ except for
$J_0=-L_0, J_1=-L_1, J_2=-L_2$. The dynamical evolution of
$|{q}^{[3]}_{rw}|^{2}$(triangular pattern) for the parametric choice
$K_0=1, J_0=-L_0, J_1=L_1, J_2=-L_2, L_0=0, L_1=500, L_2=0, a=1,
c=-1$ is plotted in the Figure. 9 and its density map in Figure. 10.
Similarly, there is another kind of third-order rogue wave, for
example, $|{q}^{[3]}_{rw}|^{2}$ is higher than third-rogue above.
The dynamical evolution of $|{q}^{[3]}_{rw}|^{2}$(fundament pattern)
for the parametric choice $J_0=-L_0, J_1=-L_1, J_2=-L_2, a=1, c=-1$
is plotted in the Figure. 11. Similarly, the density plot of Figure.
11 and the corresponding density plot in Figure. 12. Note that
$|{q}^{[3]}_{rw}|^{2}$ has only two free parameters $a$ and $c$
under the condition $J_0=-L_0, J_1=-L_1, J_2=-L_2$.

\noindent {\bf Case 3).}  When $n=4$, substituting
eq.(\ref{eigenfunfornonzeroseed}) into eq.(\ref{rntt4}) one can generate the
four-order rogue waves with eleven free parameters. Note that under
the condition $J_3>>J_i, L_3>>L_i (i=0,1,2)$  or $J_2>>J_i, L_2>>L_i
(i=0,1,3)$ or $J_1>>J_i, L_1>>L_i (i=0,2,3)$
 except for $J_0=-L_0, J_1=-L_1,
J_2=-L_2,J_3=-L_3$, the four-rogue waves can split into ten
first-order rogue wave rather. One kind of circular rogue wave
\cite{Circular} may be constructed by the condition $J_3>>J_i,
L_3>>L_i (i=0,1,2)$ except for $J_0=-L_0, J_1=-L_1,
J_2=-L_2,J_3=-L_3$. The dynamical evolution of
$|{q}^{[4]}_{rw}|^{2}$, which is a fourth order rouge wave
consisting of a ring structure (outer,seven peaks) and a fundamental
pattern of the second rogue wave (inner), for the parametric choice
$K_0=1, J_0=-L_0, J_1=-L_1, J_2=-L_2, J_3=L_3, L_0=L_1=L_2=5,
L_3=2000, a=1, c=-1$ is plotted in the Figure. 13 and the
corresponding density plot in Figure. 14. In addition, the dynamical evolution of
$|{q}^{[4]}_{rw}|^{2}$, which is another circular pattern rogue wave
\cite{Circular}, for the parametric choice $K_0=1, J_0=L_0,
J_1=-L_1, J_2=-L_2, J_3=L_3, L_0=50, L_1=10, L_2=0, L_3=2000, a=1,
c=-1$ is plotted in the Figure. 15. and its density map is portrayed
in  Figure.16. Note that the inner structure of Figure 15 (or 16) is
a triangular pattern of a second order rogue wave.  At the same
time,  the triangular pattern of the rogue wave  may be constructed
by the condition $J_1>>J_i, L_1>>L_i (i=0,2,3)$ except for
$J_0=-L_0, J_1=-L_1, J_2=-L_2,J_3=-L_3$. The dynamical evolution of
$|{q}^{[4]}_{rw}|^{2}$(triangular pattern) for the parametric choice
$K_0=1, J_0=-L_0, J_1=L_1, J_2=-L_2, J_3=-L_3, L_0=0, L_1=2000,
L_2=L_3=0, a=1, c=-1$ is plotted in the Figure. 17 and its density
plot in Figure. 18.  A pentagon pattern of the rogue wave may be
constructed by the condition $J_2>>J_i, L_2>>L_i (i=0,1,3)$ except
for $J_0=-L_0, J_1=-L_1, J_2=-L_2,J_3=-L_3$. The dynamical evolution
of $|{q}^{[4]}_{rw}|^{2}$ (pentagon pattern) for the parametric
choice $K_0=1, J_0=-L_0, J_1=-L_1, J_2=L_2, J_3=-L_3, L_0=L_1=0,
L_2=2000, L_3=0, a=1, c=-1$ is plotted in the Figure. 19. Similarly,
the density plot of Figure. 19 is correspondingly shown in Figure.
20. Similarly, there is another kind four-order rogue wave, for
example, $|{q}^{[4]}_{rw}|^{2}$ is higher than four-rogue above. The
dynamical evolution of $|{q}^{[4]}_{rw}|^{2}$ (fundamental pattern)
for the parametric choice $J_0=-L_0, J_1=-L_1, J_2=-L_2,J_3=-L_3,
a=1, c=-1$ is plotted in the Figure. 21 and the corresponding
density plot is portrayed in Figure. 22. Note that
$|{q}^{[4]}_{rw}|^{2}$ only has two free parameters $a$ and $c$
under the condition $J_0=-L_0, J_1=-L_1, J_2=-L_2,J_3=-L_3$.
According to the above analysis, the n-order rogue waves may be
controlled by $2n+3$ free parameters.

\section{Conclusions}

Thus, in this paper, considering FL system of equation which
describes nonlinear pulse propagation through single mode optical
fiber, the determinant representation of the N-fold DT for the FL
system is given in eqs.(\ref{fss1}).   The n-order rogue wave in
eq.(\ref{rntt4}) of the FL equation is obtained by this
representation from a periodic seed under the reduction condition
eq.(\ref{2nfoldredu}).  Several interesting patterns of the rogue
wave are plotted by choosing suitable parameters.

Our results provide an alternative possibility to observe rogue
waves in optical system. Moreover, from the one-fold DT, it is
interesting to observe that the DT of the FL system exhibits the
following novelty in comparison with other integrable models like
the AKNS and the KN systems: the DT matrix of the FL system has
three different terms depending on $\lambda$. Thus, the DT as well
as the rogue wave of the FL system holds novel features, in
comparison with the DT and rogue wave solutions of the standard
integrable counterparts like the AKNS and the KN systems. The
n-order  rogue waves of the FL equation are constructed by using the
determinant representation of the DT with $2n+3$ free parameters.

{\bf Acknowledgments} {\noindent \small This work is supported by
the NSF of China under Grant No.10971109 and K.C.Wong Magna Fund in
Ningbo University. Jingsong He is also supported by Program for NCET
under Grant No.NCET-08-0515 and Natural Science Foundation of Ningbo
under Grant No.2011A610179. We thank Prof. Yishen Li(USTC,Hefei,
China) for his useful suggestions on the rogue wave. J. He thank
Prof. A.S.Fokas (Cambridge University) for many helps on this
paper.}


\begin{figure}[ht]
\setlength{\unitlength}{0.1cm}
\begin{minipage}[t]{8cm}
\epsfig{file=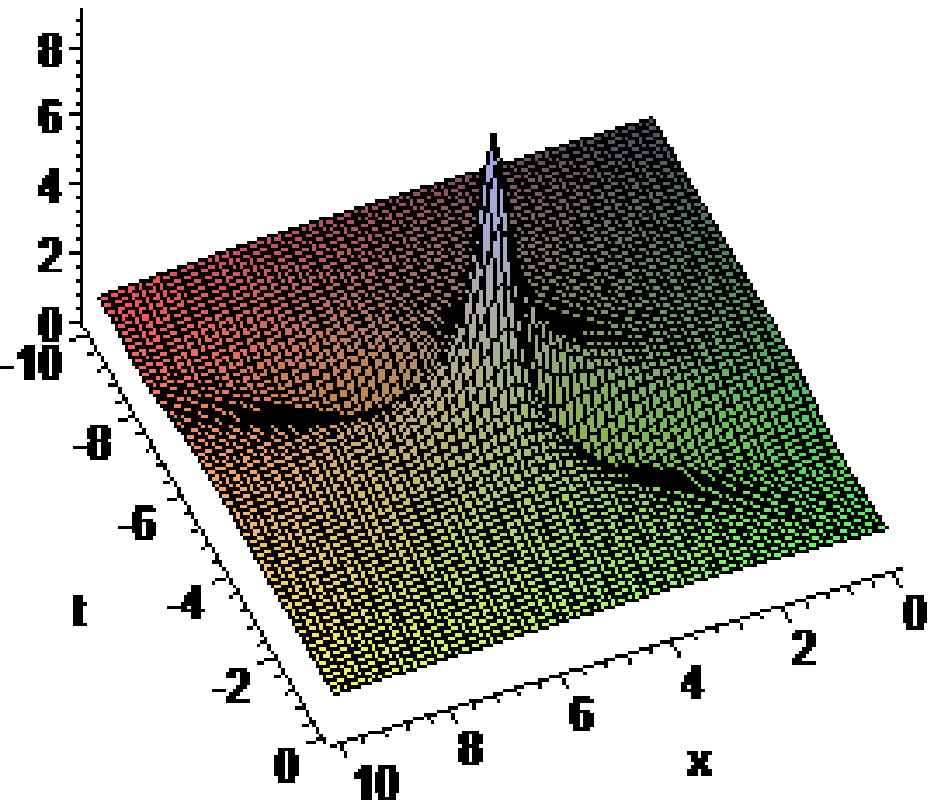,width=5cm}\vspace{0cm}\caption{{The
dynamical evolution of $|{q}^{[1]}_{rw}|^{2}$ with specific
parameters $K_0=1, J_0=L_0=10, a=1, c=-1 $.}}
\end{minipage}
\hspace{0.5cm} 
\begin{minipage}[t]{8cm}
\epsfig{file=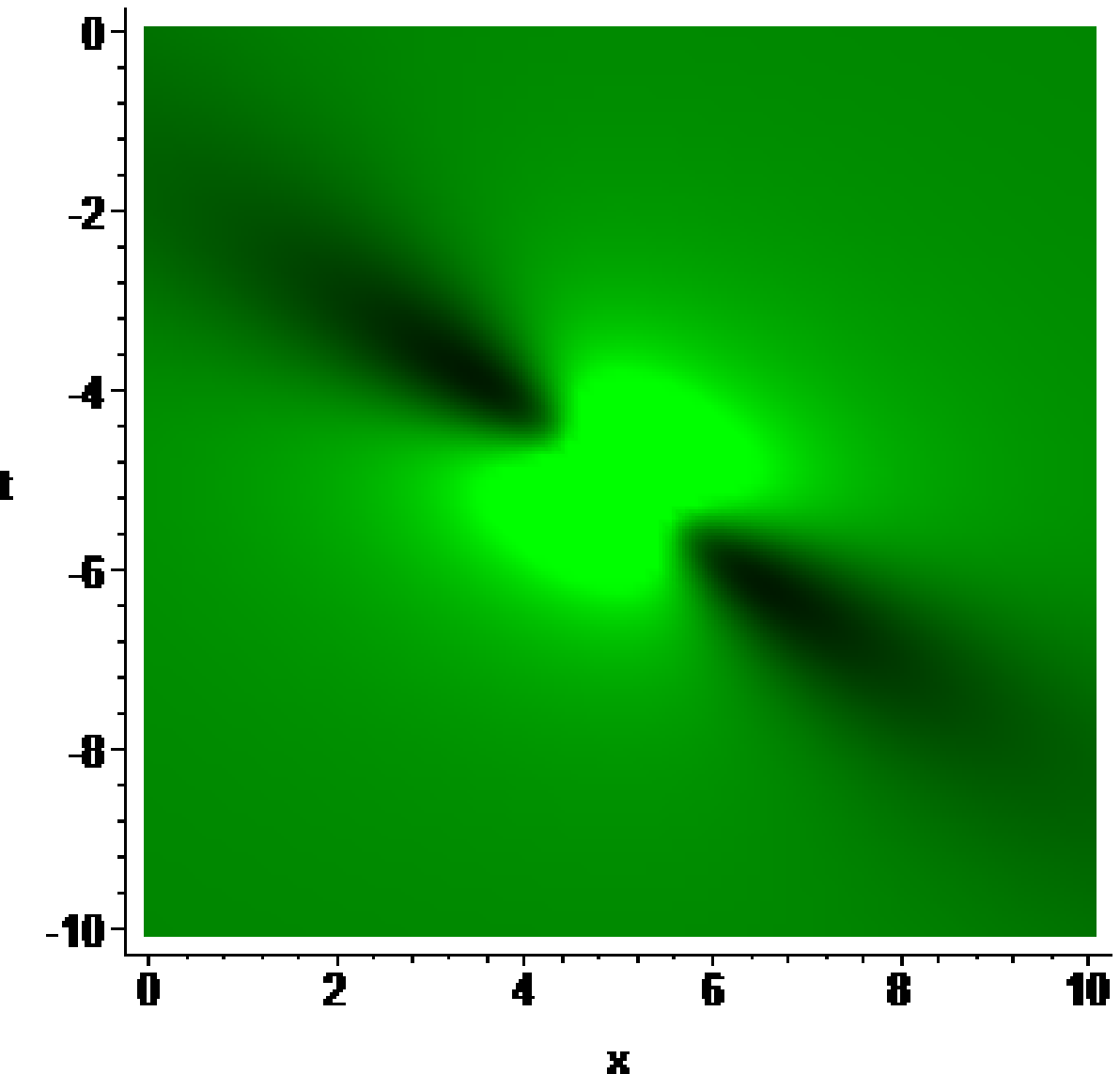,width=5cm}\vspace{0cm}\caption{{Contour
plot of the wave amplitudes of $|{q}^{[1]}_{rw}|^{2}$ for the values
used in Figure 1.}}
\end{minipage}
\end{figure}
\begin{figure}[ht]
\setlength{\unitlength}{0.1cm}
\begin{minipage}[t]{8cm}
\epsfig{file=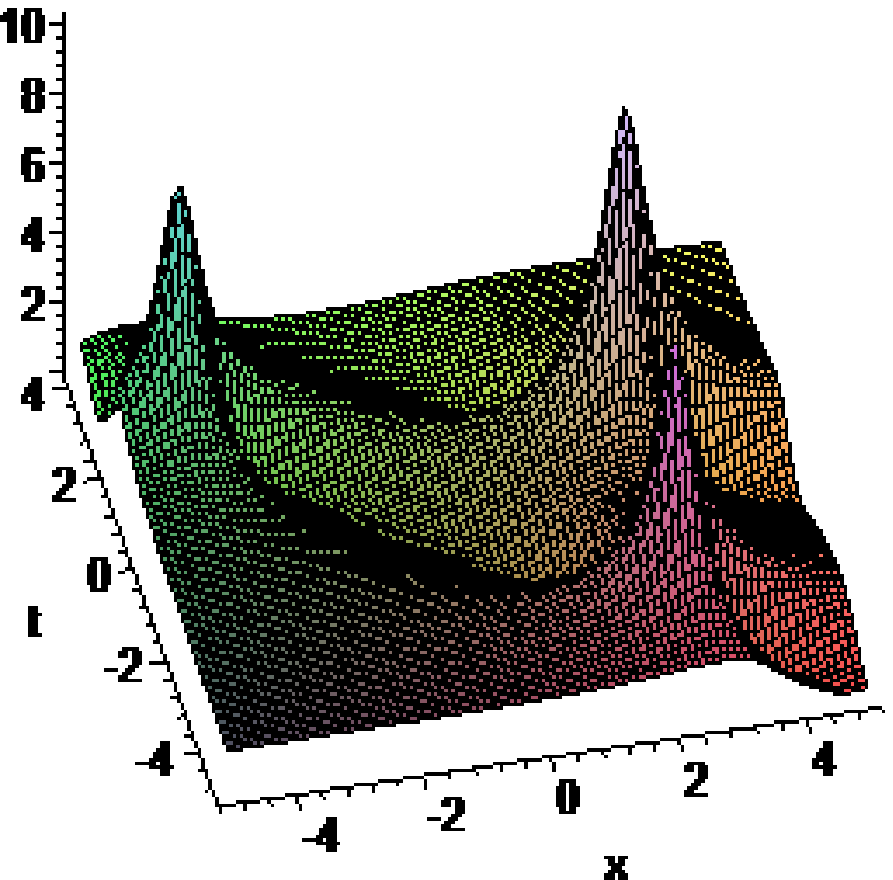,width=5cm}\vspace{0cm}\caption{{The
dynamical evolution of $|{q}^{[2]}_{rw}|^{2}$ with specific
parameters $K_0=1, J_0=-L_0, J_1=L_1, L_0=-5, L_1=20, a=1, c=-1$.}}
\end{minipage}
\hspace{0.5cm} 
\begin{minipage}[t]{8cm}
\epsfig{file=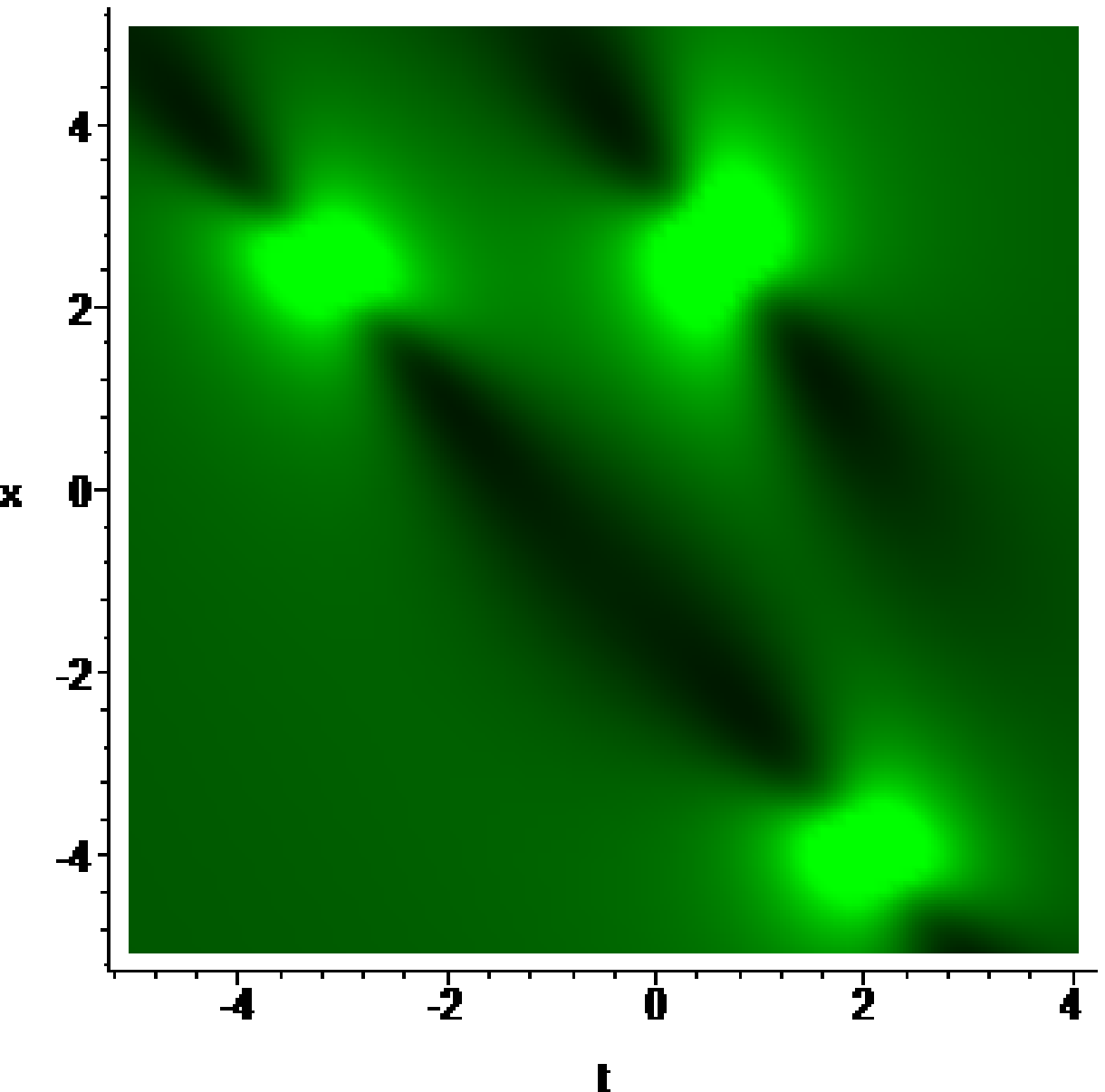,width=5cm}\vspace{0cm}\caption{{Contour
plot of the wave amplitudes of $|{q}^{[2]}_{rw}|^{2}$ for the values
used in Figure 3.}}
\end{minipage}
\end{figure}
\begin{figure}[ht]
\setlength{\unitlength}{0.1cm}
\begin{minipage}[t]{8cm}
\epsfig{file=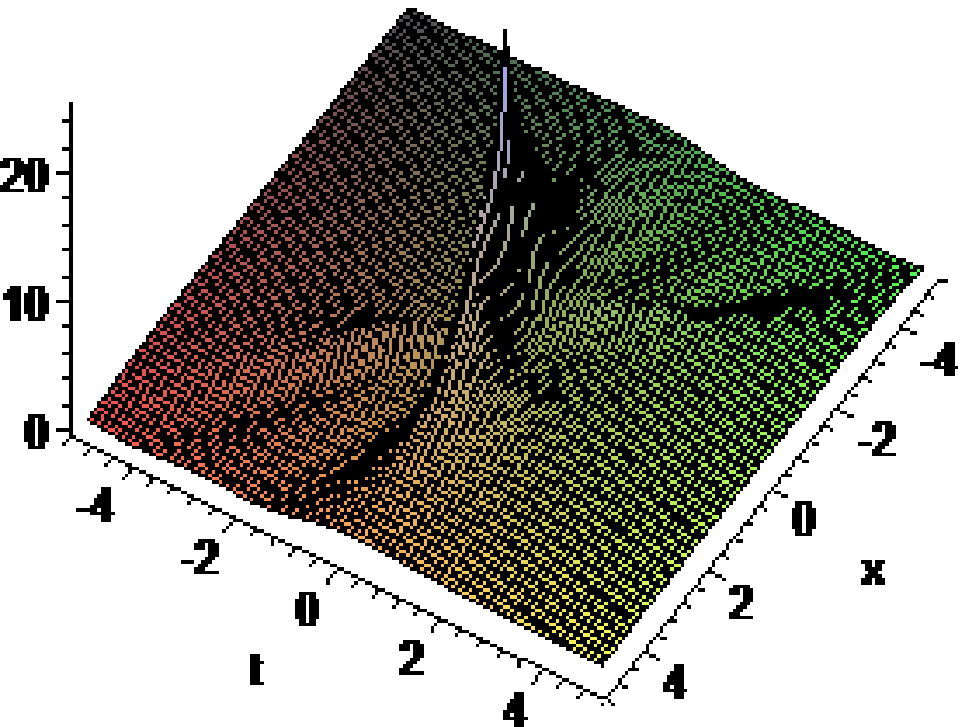,width=5cm}\vspace{0cm}\caption{{The
dynamical evolution of $|{q}^{[2]}_{rw}|^{2}$ with specific
parameters $J_0=-L_0, J_1=-L_1, a=1, c=-1$.}}
\end{minipage}
\hspace{0.5cm} 
\begin{minipage}[t]{8cm}
\epsfig{file=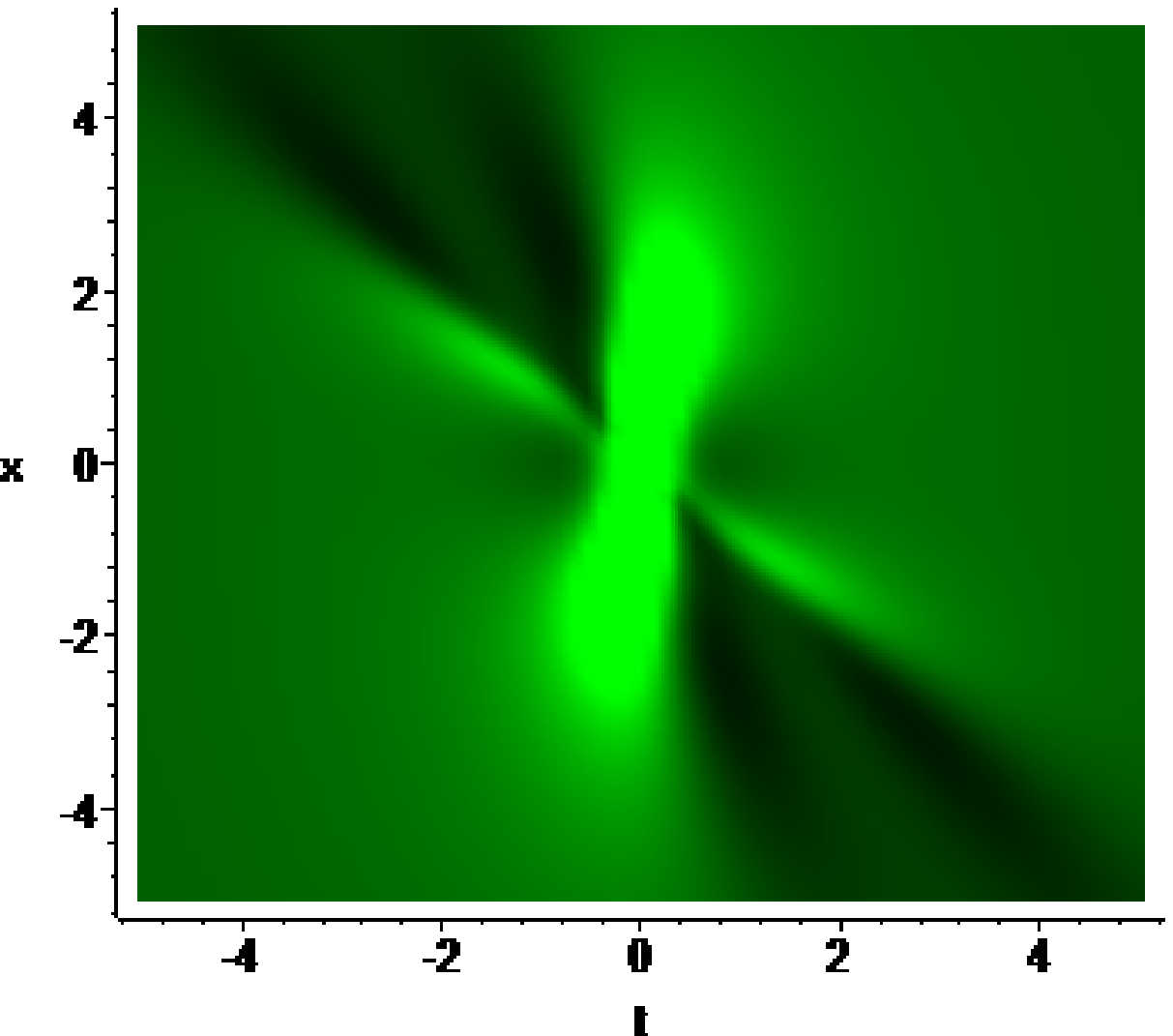,width=5cm}\vspace{0cm}\caption{{Contour
plot of the wave amplitudes of $|{q}^{[2]}_{rw}|^{2}$ for the values
used in Figure 5.}}
\end{minipage}
\end{figure}

\begin{figure}[ht]
\setlength{\unitlength}{0.1cm}
\begin{minipage}[t]{8cm}
\epsfig{file=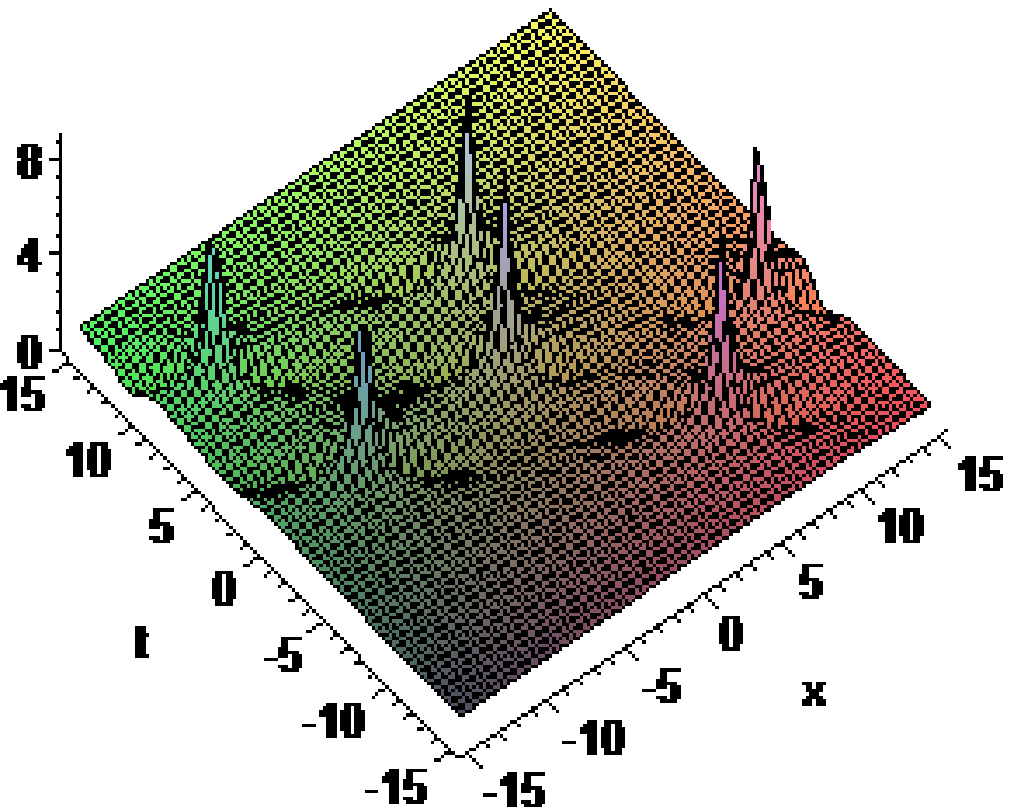,width=5cm}\vspace{0cm}\caption{{The
dynamical evolution of $|{q}^{[3]}_{rw}|^{2}$ with specific
parameters $K_0=1, J_0=-L_0, J_1=-L_1, J_2=L_2, L_0=5, L_1=5,
L_2=3000, a=1, c=-1$.}}
\end{minipage}
\hspace{0.5cm} 
\begin{minipage}[t]{8cm}
\epsfig{file=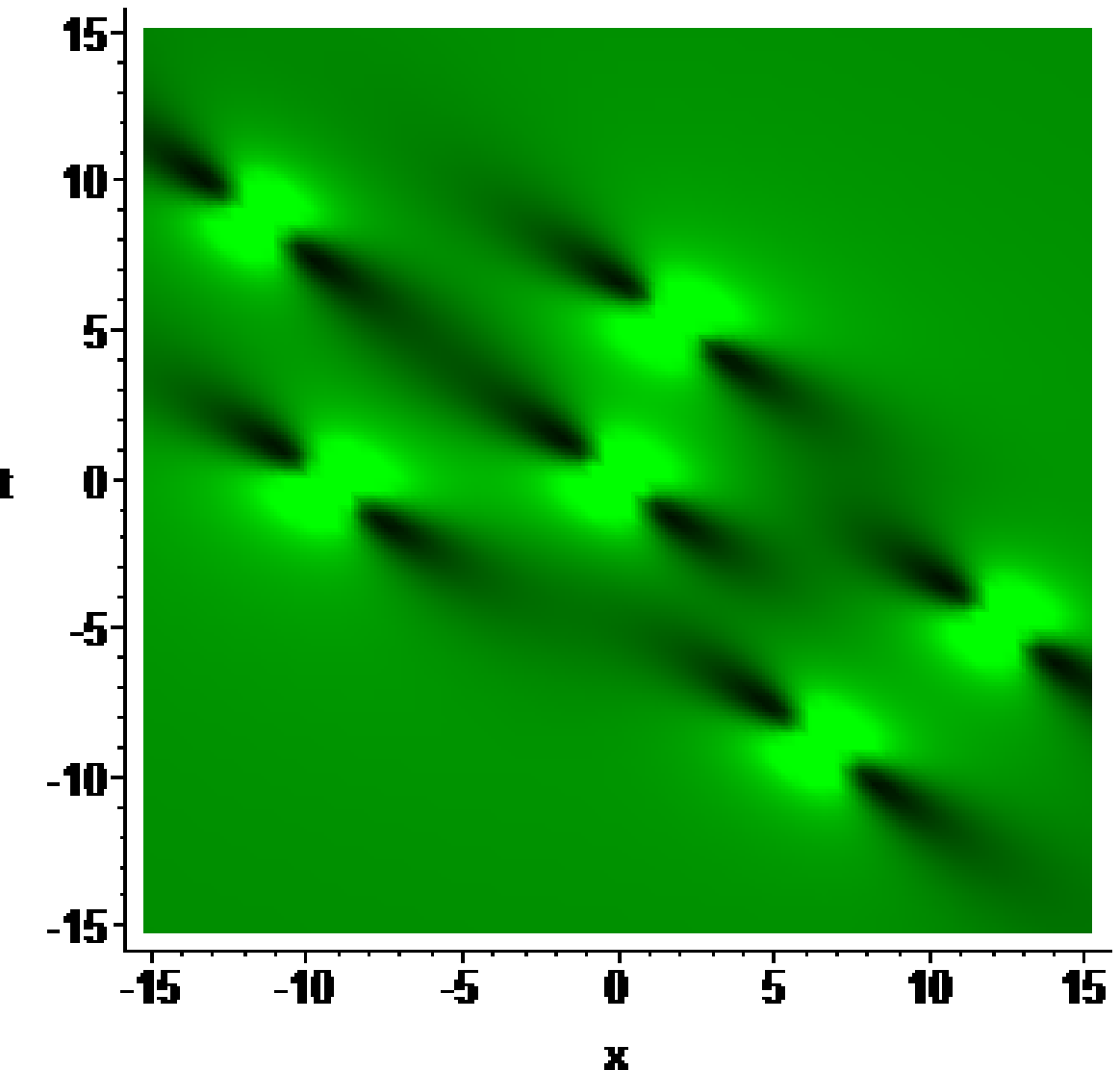,width=5cm}\vspace{0cm}\caption{{Contour
plot of the wave amplitudes of $|{q}^{[3]}_{rw}|^{2}$ for the values
used in Figure 7.}}
\end{minipage}
\end{figure}
\begin{figure}[ht]
\setlength{\unitlength}{0.1cm}
\begin{minipage}[t]{8cm}
\epsfig{file=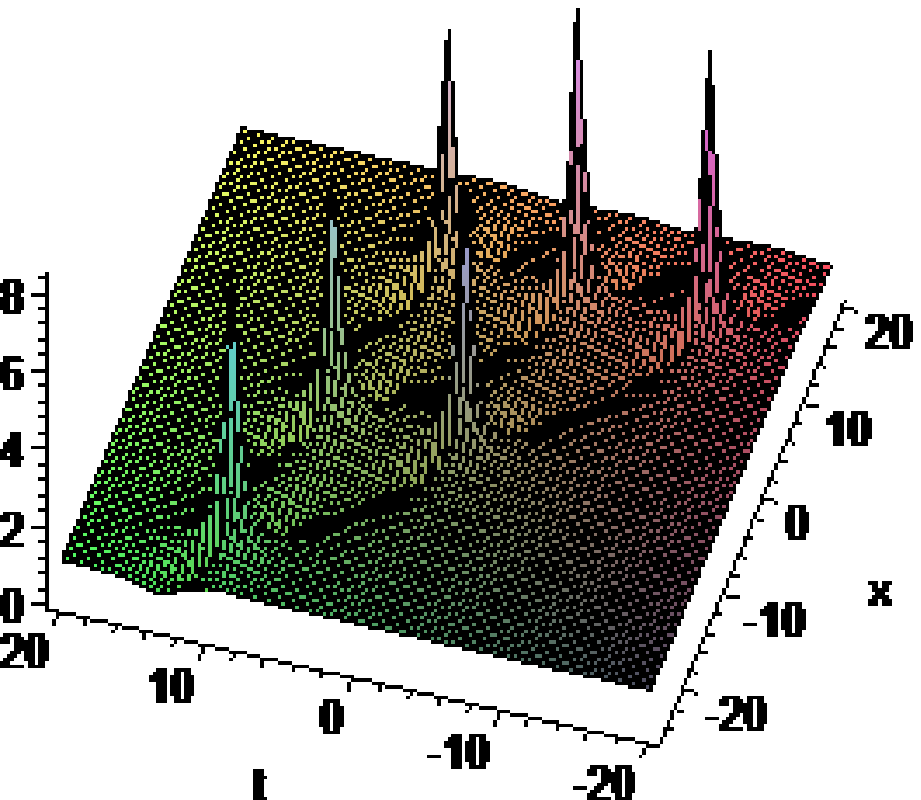,width=5cm}\vspace{0cm}\caption{{The
dynamical evolution of $|{q}^{[3]}_{rw}|^{2}$ with specific
parameters $K_0=1, J_0=-L_0, J_1=L_1, J_2=-L_2, L_0=0, L_1=500,
L_2=0, a=1, c=-1$.}}
\end{minipage}
\hspace{0.5cm} 
\begin{minipage}[t]{8cm}
\epsfig{file=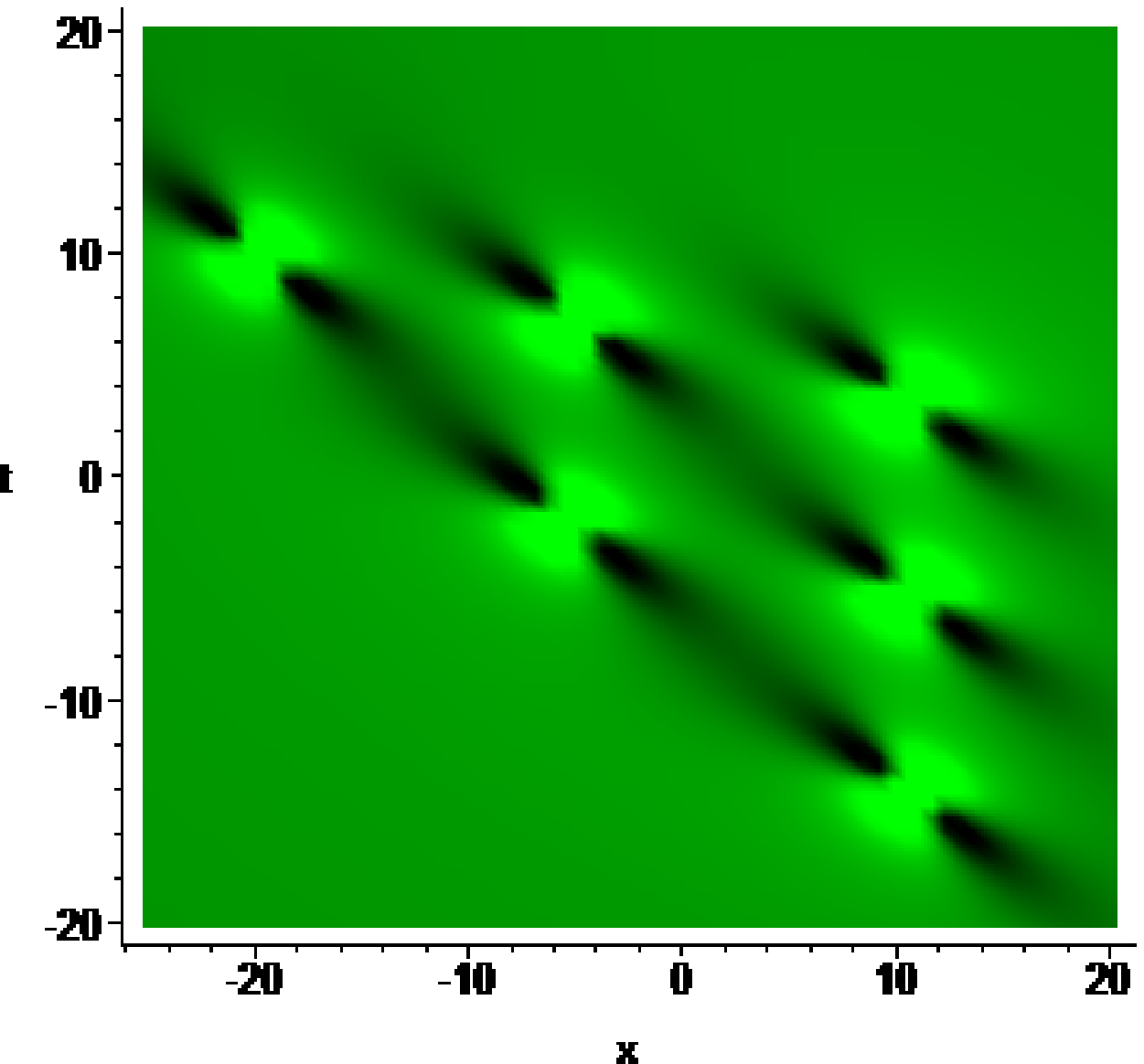,width=5cm}\vspace{0cm}\caption{{Contour
plot of the wave amplitudes of $|{q}^{[3]}_{rw}|^{2}$ for the values
used in Figure 9.}}
\end{minipage}
\end{figure}
\begin{figure}[ht]
\setlength{\unitlength}{0.1cm}
\begin{minipage}[t]{8cm}
\epsfig{file=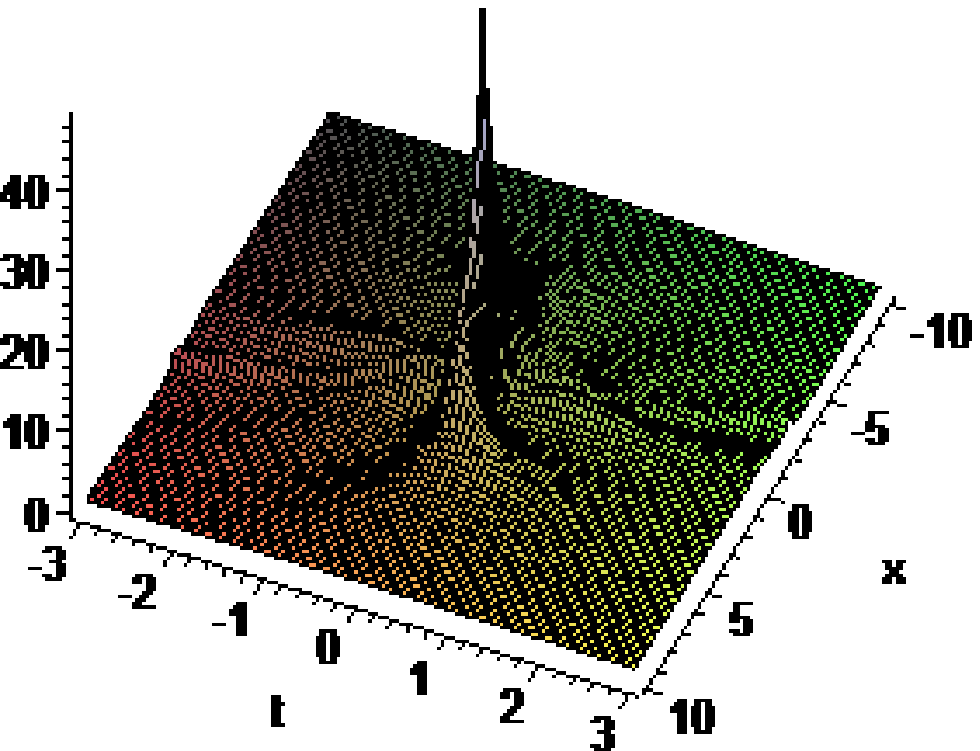,width=5cm}\vspace{0cm}\caption{{The
dynamical evolution of $|{q}^{[3]}_{rw}|^{2}$ with specific
parameters $J_0=-L_0, J_1=-L_1, J_2=-L_2, a=1, c=-1$.}}
\end{minipage}
\hspace{0.5cm} 
\begin{minipage}[t]{8cm}
\epsfig{file=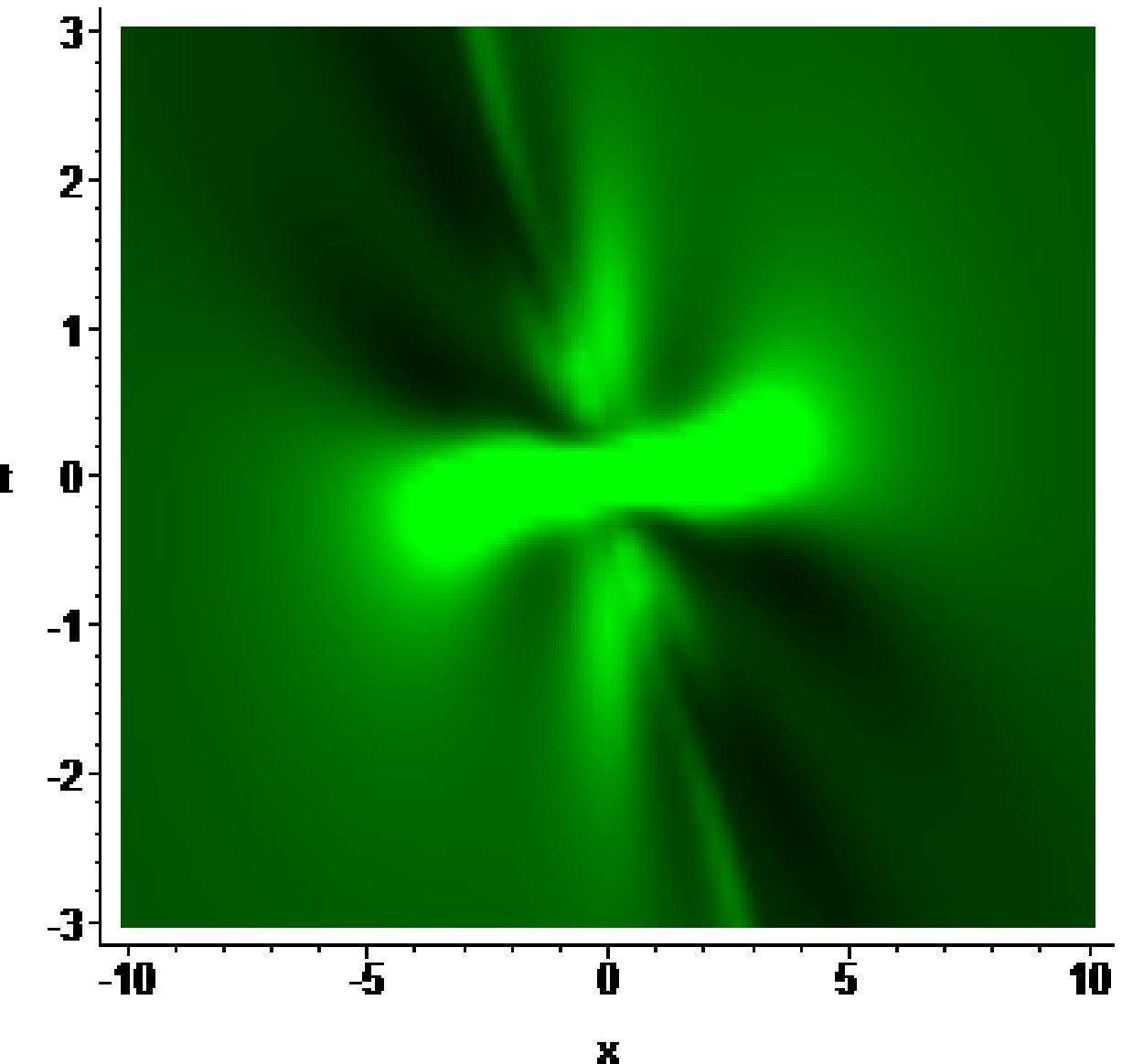,width=5cm}\vspace{0cm}\caption{{Contour
plot of the wave amplitudes of $|{q}^{[3]}_{rw}|^{2}$ for the values
used in Figure 11.}}
\end{minipage}
\end{figure}

\begin{figure}[ht]
\setlength{\unitlength}{0.1cm}
\begin{minipage}[t]{8cm}
\epsfig{file=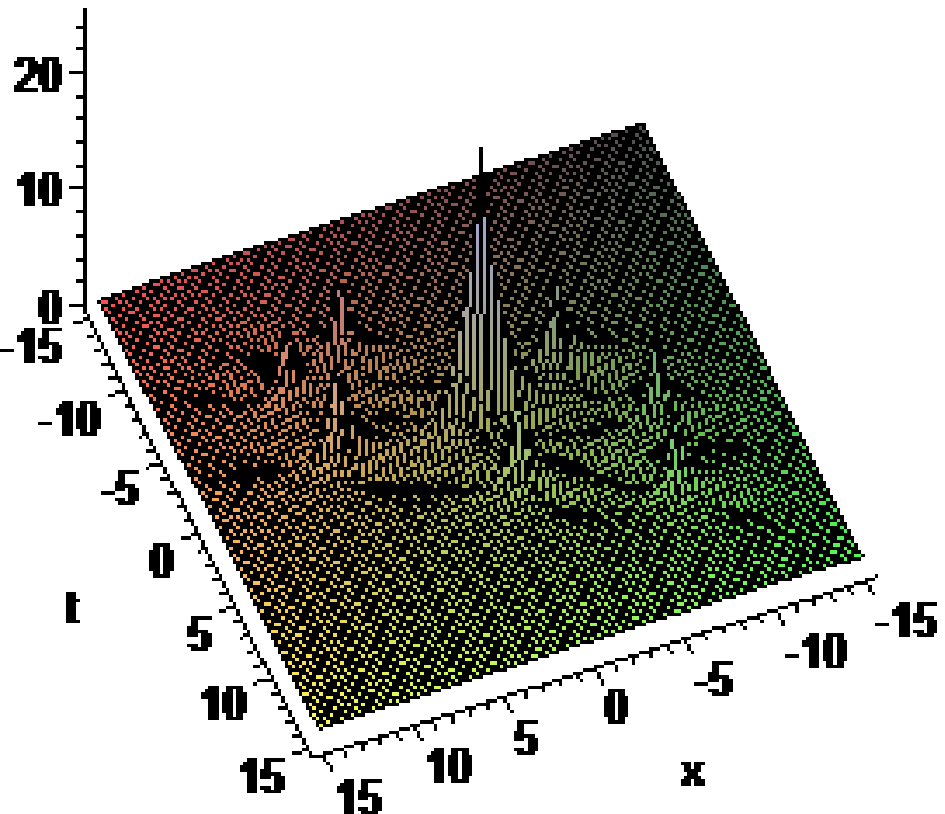,width=5cm}\vspace{0cm}\caption{{The
dynamical evolution of $|{q}^{[4]}_{rw}|^{2}$ with specific
parameters $K_0=1, J_0=-L_0, J_1=-L_1, J_2=-L_2, J_3=L_3,
L_0=L_1=L_2=5, L_3=2000, a=1, c=-1$.}}
\end{minipage}
\hspace{0.5cm} 
\begin{minipage}[t]{8cm}
\epsfig{file=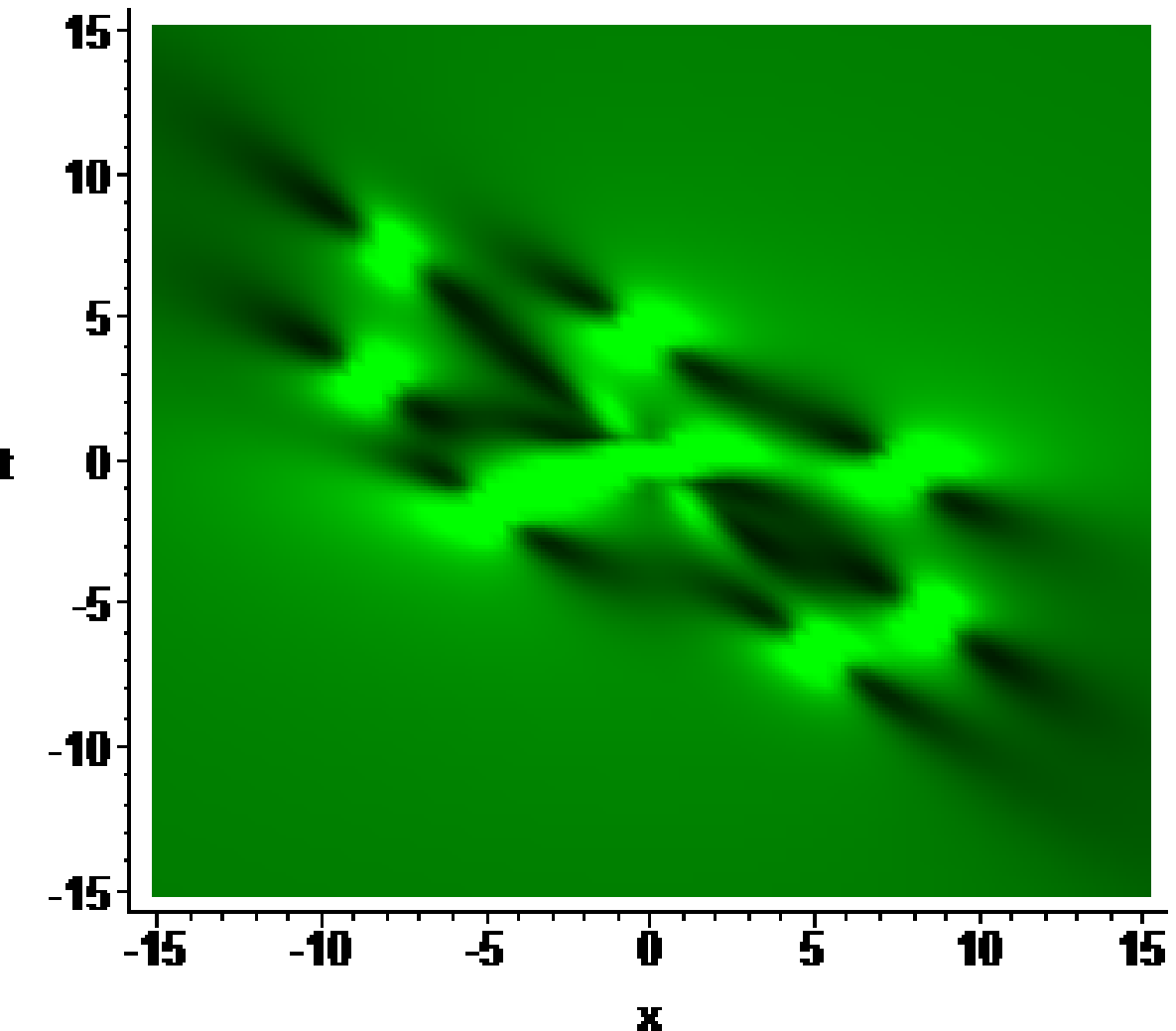,width=5cm}\vspace{0cm}\caption{{Contour
plot of the wave amplitudes of $|{q}^{[4]}_{rw}|^{2}$ for the values
used in Figure 13.}}
\end{minipage}
\end{figure}
\begin{figure}[ht]
\setlength{\unitlength}{0.1cm}
\begin{minipage}[t]{8cm}
\epsfig{file=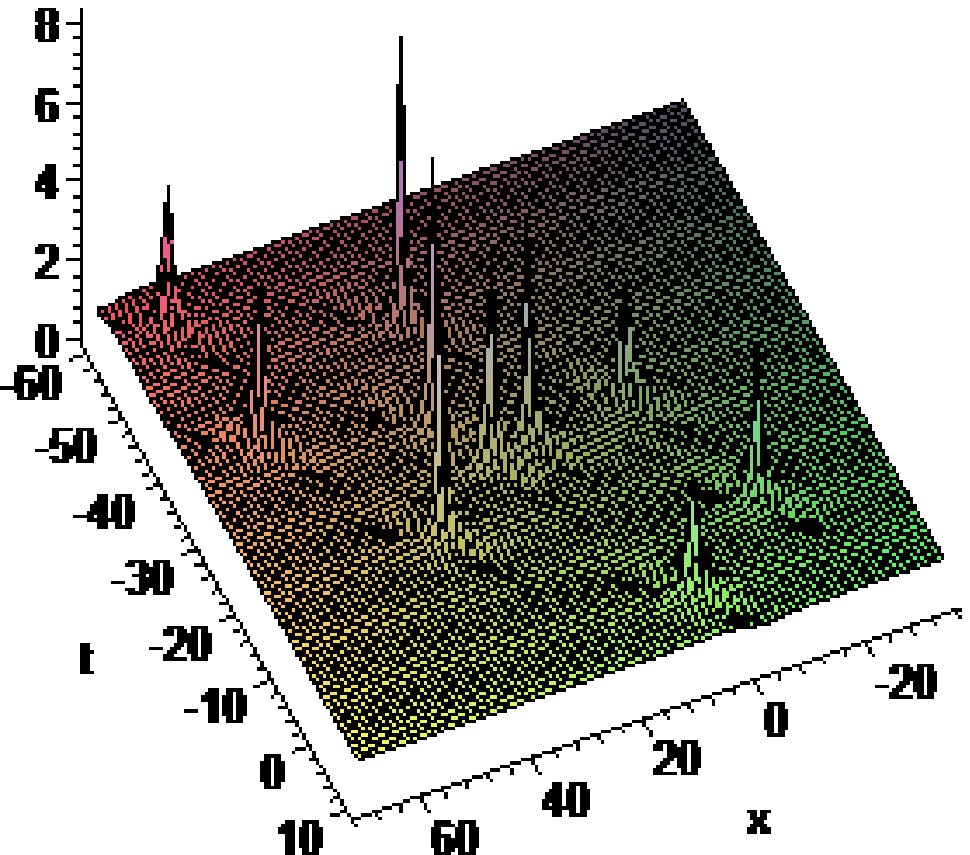,width=5cm}\vspace{0cm}\caption{{The
dynamical evolution of $|{q}^{[4]}_{rw}|^{2}$ with specific
parameters $K_0=1, J_0=L_0, J_1=-L_1, J_2=-L_2, J_3=L_3, L_0=50,
L_1=10, L_2=0, L_3=2000, a=1, c=-1$.}}
\end{minipage}
\hspace{0.5cm} 
\begin{minipage}[t]{8cm}
\epsfig{file=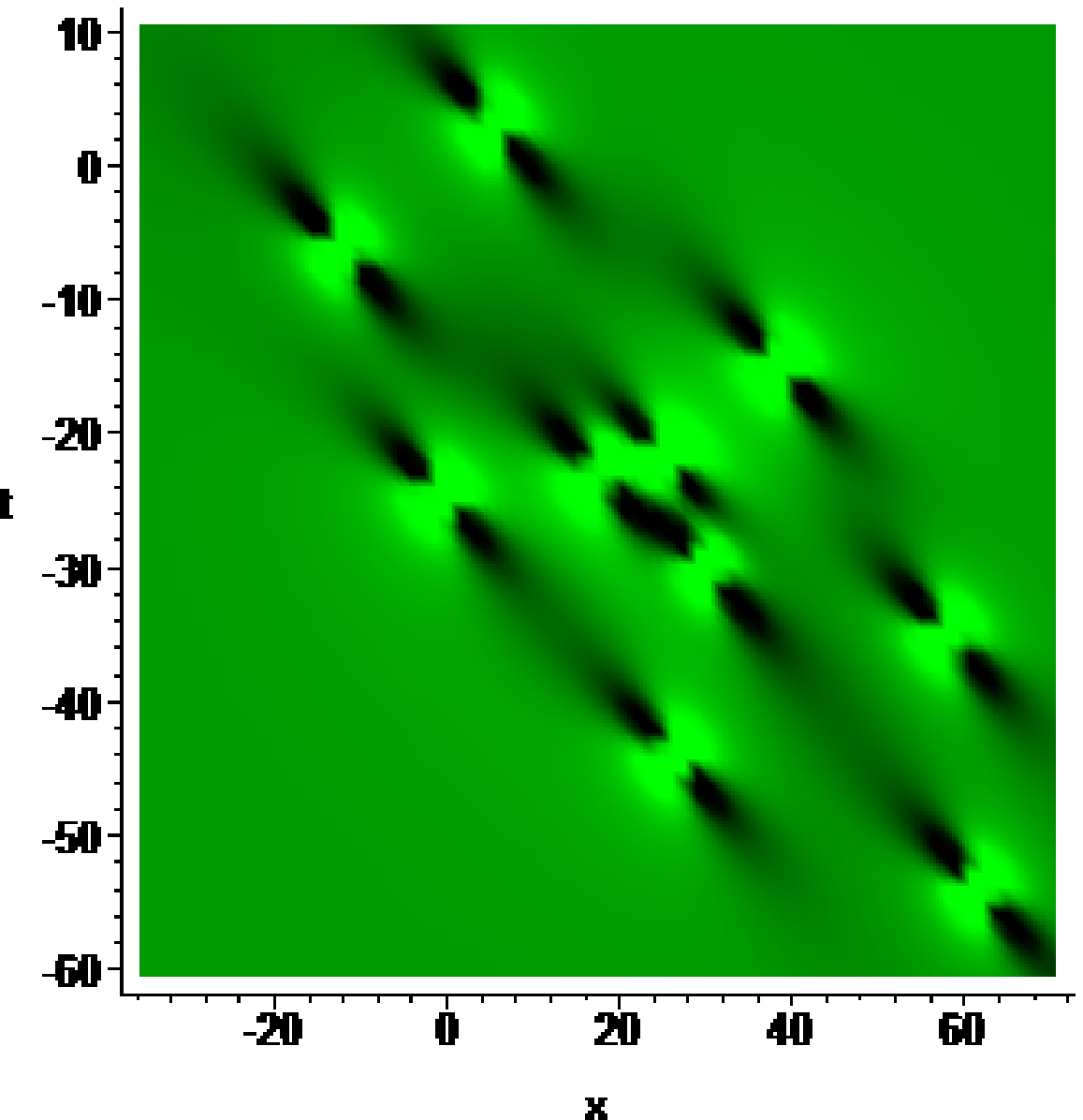,width=5cm}\vspace{0cm}\caption{{Contour
plot of the wave amplitudes of $|{q}^{[4]}_{rw}|^{2}$ for the values
used in Figure 15.}}
\end{minipage}
\end{figure}
\begin{figure}[ht]
\setlength{\unitlength}{0.1cm}
\begin{minipage}[t]{8cm}
\epsfig{file=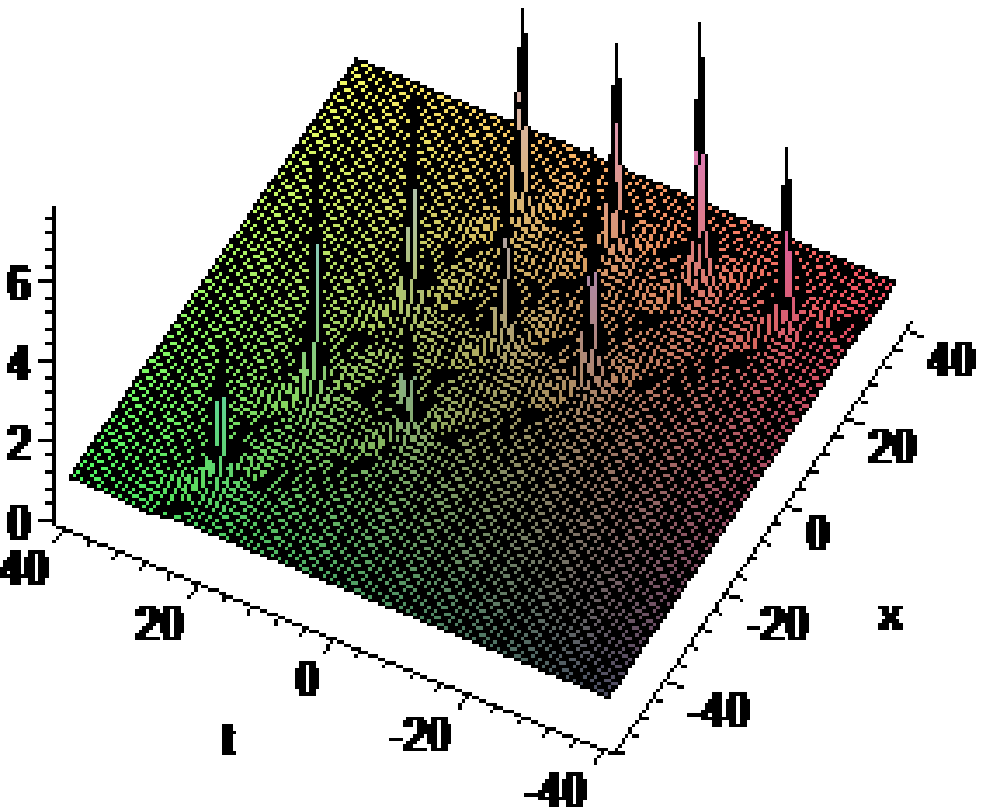,width=5cm}\vspace{0cm}\caption{{The
dynamical evolution of $|{q}^{[4]}_{rw}|^{2}$ with specific
parameters $K_0=1, J_0=-L_0, J_1=L_1, J_2=-L_2, J_3=-L_3, L_0=0,
L_1=2000, L_2=L_3=0, a=1, c=-1$.}}
\end{minipage}
\hspace{0.5cm} 
\begin{minipage}[t]{8cm}
\epsfig{file=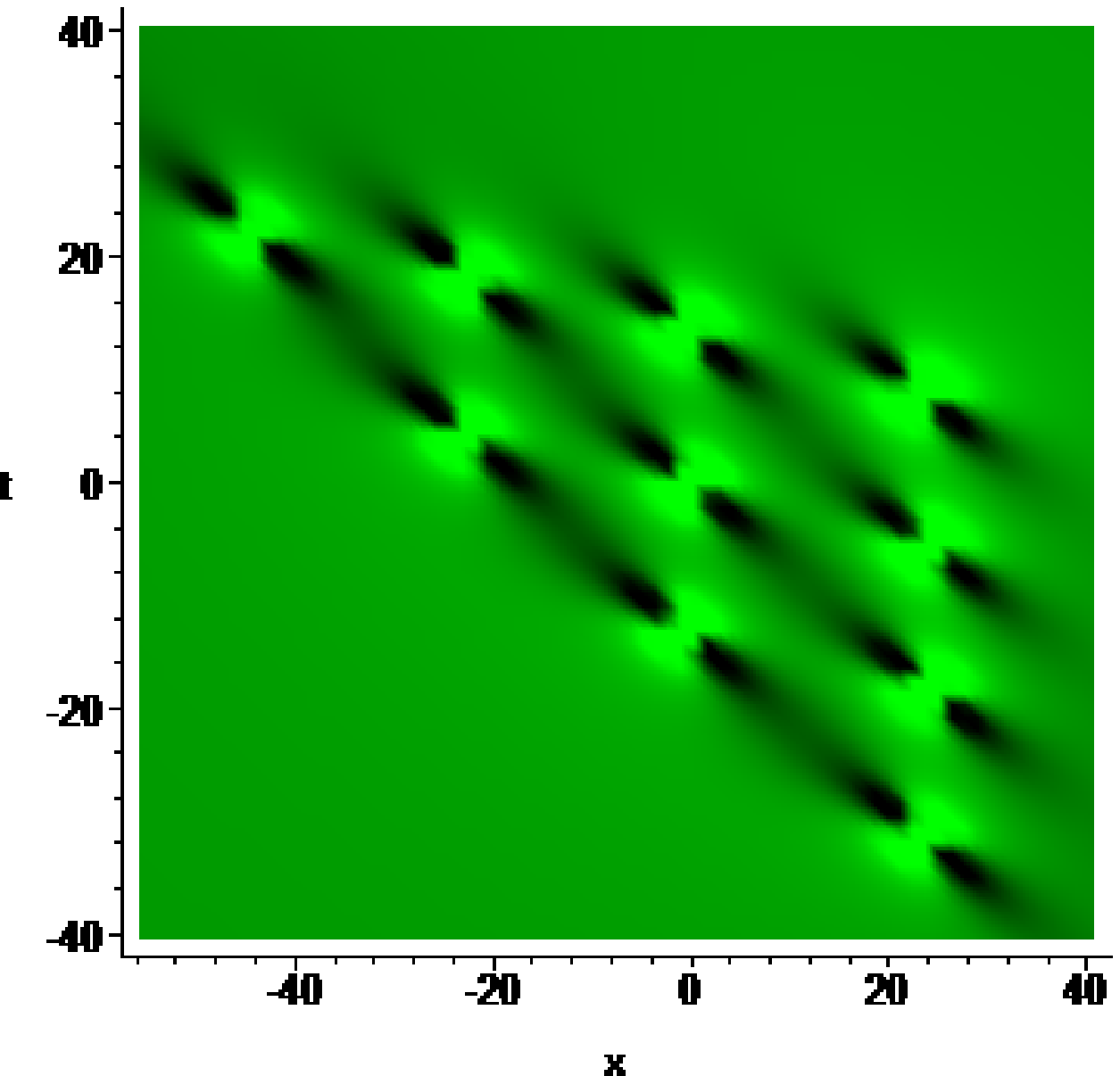,width=5cm}\vspace{0cm}\caption{{Contour
plot of the wave amplitudes of $|{q}^{[4]}_{rw}|^{2}$ for the values
used in Figure 17.}}
\end{minipage}
\end{figure}
\begin{figure}[ht]
\setlength{\unitlength}{0.1cm}
\begin{minipage}[t]{8cm}
\epsfig{file=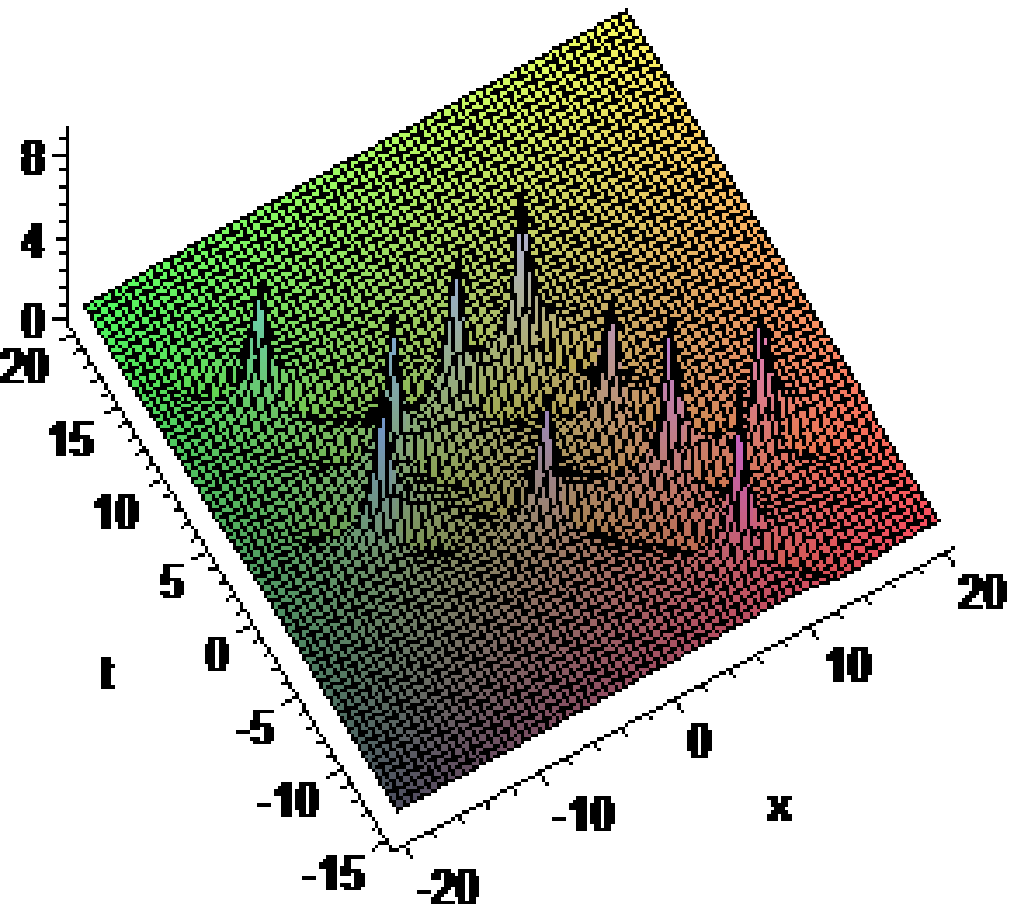,width=5cm}\vspace{0cm}\caption{{The
dynamical evolution of $|{q}^{[4]}_{rw}|^{2}$ with specific
parameters $K_0=1, J_0=-L_0, J_1=-L_1, J_2=L_2, J_3=-L_3, L_0=L_1=0,
L_2=2000, L_3=0, a=1, c=-1$.}}
\end{minipage}
\hspace{0.5cm} 
\begin{minipage}[t]{8cm}
\epsfig{file=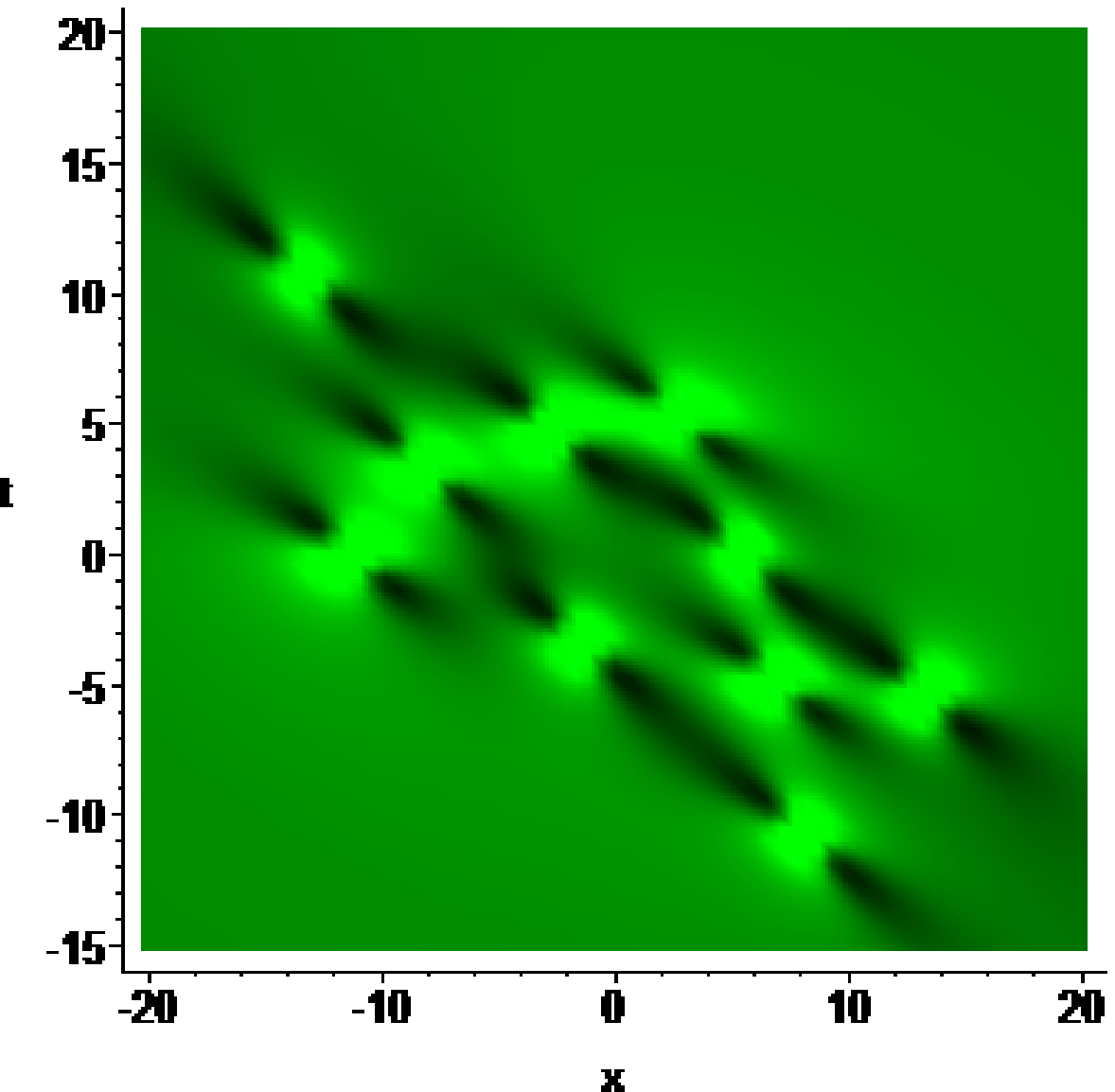,width=5cm}\vspace{0cm}\caption{{Contour
plot of the wave amplitudes of $|{q}^{[4]}_{rw}|^{2}$ for the values
used in Figure 19.}}
\end{minipage}
\end{figure}
\begin{figure}[ht]
\setlength{\unitlength}{0.1cm}
\begin{minipage}[t]{8cm}
\epsfig{file=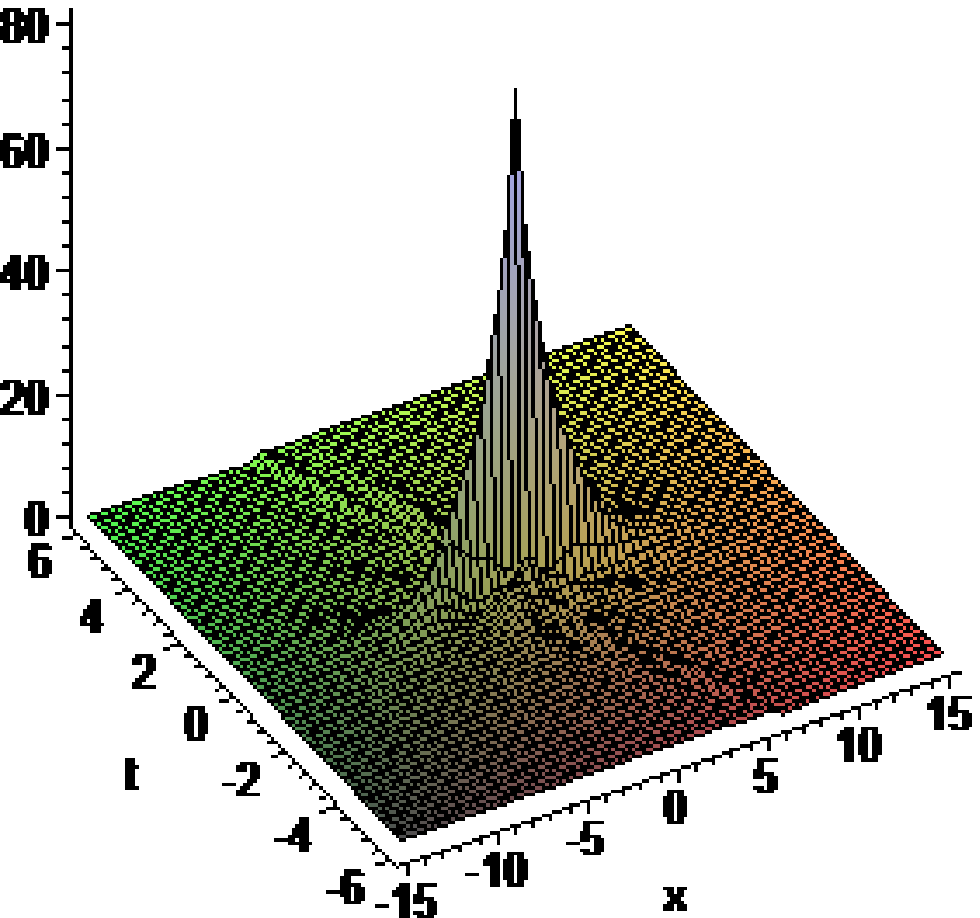,width=5cm}\vspace{0cm}\caption{{The
dynamical evolution of $|{q}^{[4]}_{rw}|^{2}$ with specific
parameters $J_0=-L_0, J_1=-L_1, J_2=-L_2, J_3=-L_3, a=1, c=-1$.}}
\end{minipage}
\hspace{0.5cm} 
\begin{minipage}[t]{8cm}
\epsfig{file=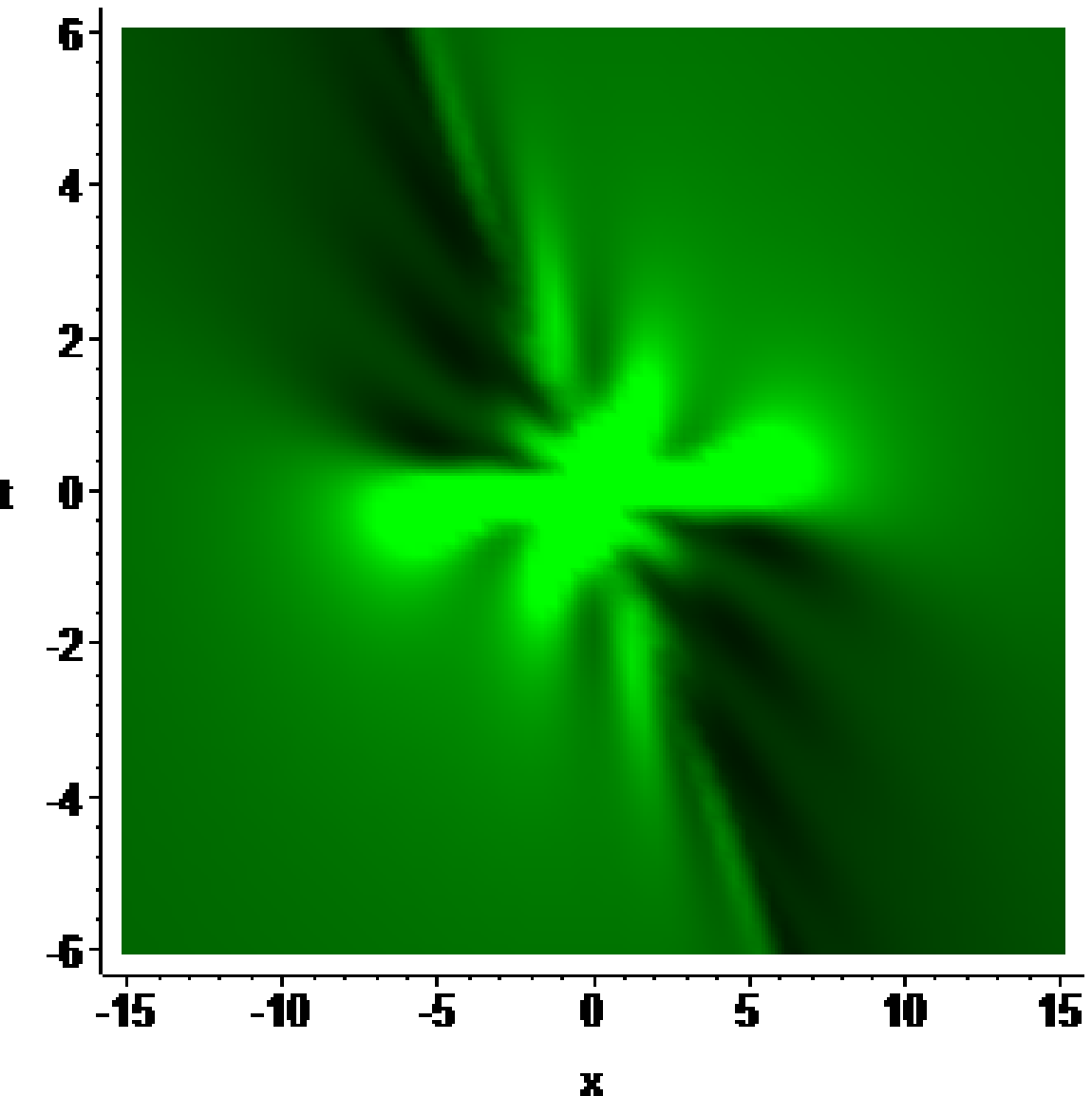,width=5cm}\vspace{0cm}\caption{{Contour
plot of the wave amplitudes of $|{q}^{[4]}_{rw}|^{2}$ for the values
used in Figure 21.}}
\end{minipage}
\end{figure}


\end{document}